\documentclass[prb,aps,amssymb,twocolumn,superscriptaddress,notitlepage]{revtex4-1}
\usepackage{amsmath}
\usepackage{amssymb}
\usepackage{amsthm}
\usepackage{amsfonts}
\usepackage{listings}
\usepackage{enumerate}
\usepackage{latexsym}
\usepackage{psfrag}
\usepackage{bm}
\usepackage{graphicx}
\usepackage[caption=false]{subfig}
\usepackage{blkarray}
\usepackage{array}
\usepackage{color}
\usepackage[normalem]{ulem}
\usepackage{hyperref}

\usepackage{tikz}
\usepackage[capitalise]{cleveref} 

\def\sgtwelve{Pn\bar{3}n1'}
\def\sgthirteen{Pm\bar{3}n1'}
\def\sgthirty{Ia\bar{3}d1'}
\def\sgseven{P4321'}
\def\sgeight{P4_2321'}

\def\dk{\Delta_\mathbf{k}}
\makeatletter
\DeclareRobustCommand{\element}[1]{\@element#1\@nil}
\def\@element#1#2\@nil{%
  #1%
  \if\relax#2\relax\else\MakeLowercase{#2}\fi}
\makeatother

\newcolumntype{L}{>{$}l<{$}}

\newtheorem*{defn*}{Definition}

\newcommand{\half}{\frac{1}{2}}

\begin{document}
\title{Multifold nodal points in magnetic materials}

\author{Jennifer Cano}
\affiliation{Department of Physics and Astronomy, Stony Brook University, Stony Brook, New York 11974, USA}
\affiliation{Center for Computational Quantum Physics, The Flatiron Institute, New York, New York 10010, USA}

\author{Barry Bradlyn}
\affiliation{Department of Physics and Institute for Condensed Matter Theory, University of Illinois at Urbana-Champaign, Urbana, IL, 61801-3080, USA}

\author{M. G. Vergniory}
\affiliation{Donostia International Physics Center, P. Manuel de Lardizabal 4, 20018 Donostia-San Sebastian, Spain}
\affiliation{IKERBASQUE, Basque Foundation for Science, Maria Diaz de Haro 3, 48013 Bilbao, Spain}

\date{\today}
\begin{abstract}
We describe the symmetry protected nodal points that can exist in magnetic space groups and show that only $3$-, $6$-, and $8$-fold degeneracies are possible (in addition to the $2$- and $4$-fold degeneracies that have already been studied.)
The $3$- and $6$-fold degeneracies are derived from ``spin-$1$'' Weyl fermions.
The $8$-fold degeneracies come in different flavors.
In particular, we distinguish between $8$-fold fermions that realize nonchiral ``Rarita-Schwinger fermions'' and those that can be described as four degenerate Weyl fermions.
We list the (magnetic and non-magnetic) space groups where these exotic fermions can be found. {We further show that in several cases, a magnetic translation symmetry pins the Hamiltonian of the multifold fermion to an idealized exactly solvable point that is not achievable in non-magnetic crystals without fine-tuning.}
Finally, we present known compounds that may host these fermions and methods for systematically finding more candidate materials.
\end{abstract}
\maketitle

\section{Introduction}\label{sec:intro}

The prediction and observation of Weyl\cite{Wan11,Weng15,Huang15,Xu15,Lv15,Xu15a,Lv15a,Xiong2015} and Dirac\cite{Young12,Wang12,Liu14,Liu14a,Steinberg14} fermions catalyzed an intense theoretical and experimental search for topological nodal semimetals in condensed matter systems.\cite{Gibson2015,Watanabe16,NaturePaper,Po2017,Vergniory18,Chen18,Tang2018,Zhang2018}
These systems provide a solid state realization of the chiral\cite{Xiong2015,Huang2015} and gravitational anomalies\cite{Lucas2016,Gooth2017} and exhibit many novel physical properties, such as gapless Fermi arc surface states,\cite{Wan11} extremely large magnetoresistance,\cite{Liang2015} and giant nonlinear optical response.\cite{Morimoto2016,Wu2017}

Topological semimetals also illustrate the interplay between symmetry and topology: Weyl fermions can exist without any symmetry, but are forbidden by the combination of time reversal and inversion symmetry.
On the other hand, Dirac fermions rely on a combination of crystal symmetries for their existence and their dispersion depends on the symmetries that protect them.\cite{Yang2014}
Crystal symmetries can also protect nodal lines.\cite{Burkov2011,Carter2012,Chiu14,Fang2015,Xie2015,Kim2015,Yu2015,Chan2016,Bian2016,Bzdusek16,Liang2016-2}

In Ref.~\onlinecite{Bradlyn2016}, we completed the catalogue of nodal point fermions that exist in systems with time-reversal symmetry and spin-orbit coupling by showing that non-symmorphic symmetries can stabilize 3-, 6-, and 8-fold degeneracies at the corners of the Brillouin zone and, furthermore, that \textit{chiral} 4-fold degeneracies are possible.
(We use chiral to refer to a nodal point that is a source/sink of Berry curvature. This should not be confused with the usage to refer to symmetry groups with only orientation-preserving operations\cite{kramersweyl,Flicker2018}, which we refer to as structural chirality.)
The multifold fermions include higher-spin analogues of Weyl and Dirac fermions, which cannot be realized in high-energy physics because they necessarily violate Poincar\'{e} symmetry. 
Recently, multifold fermions have been observed in several experimental systems.\cite{Chang2017,Schroter2018,Sanchez2018,Rao2019} 

However, the search for nodal semimetals has mostly focused on non-magnetic materials, where time reversal symmetry is present.
There are two practical reasons for this: first, hundreds of thousands of known compounds are searchable by their crystal structure, but not by their magnetic order, in the International Crystal Structure Database (ICSD).\cite{ICSD}
Second, the phase space of non-magnetic symmetries is much smaller: there are 230 non-magnetic space groups, but 1421 additional magnetic space groups.
Yet, increasingly, attention is being shifted to magnetic compounds.
Several magnetic materials have been predicted to host Weyl\cite{Nakatsuji2015,Chang2016,Nayak2016,Wang2016,Yang2017,Kuroda2017,Noky2018,Shi2018,Xu2018,ECA} and other multi-fold fermions.\cite{Wieder-magnetic,Vergniory2018,Schoop2018}
Symmetry indicators and filling constraints have also been computed for the magnetic space groups.\cite{WatanabeMagnetic}
In addition, magnetic materials are desirable for their large degree of tuneability: by varying temperature\cite{Schoop2018} or applied magnetic field\cite{Hirschberger2016,Cano2017}, potentially many different phases can be realized.

The recent experimental progress combined with the tuneability of magnetic materials motivates the study of nodal fermions in magnetic systems. In this work, we classify the degeneracy of fermionic quasiparticles that can occur in the magnetic space groups with spin-orbit coupling: as in the non-magnetic groups, we find that 3-, 6- and 8-fold degeneracies can occur, in addition to the 2- and 4-fold degeneracies already known.
We identify which of these are chiral and thus can be detected by their gapless surface states and optical response\cite{Flicker2018}.
We then describe candidate materials to realize these fermions.

As part of our investigation, we uncover space groups whose symmetries protect 8-fold Rarita-Schwinger (RS) fermions.
While ``Rarita-Schwinger Weyl'' fermions -- the chiral halves of a RS fermions -- have already been predicted to exist in condensed matter systems\cite{Bradlyn2016,Liang2016-1,Tang2017}, the full RS fermion had not been explicitly identified.
We find that these fermions can exist in both magnetic and non-magnetic space groups (the latter case was included in our classification in Ref.~\onlinecite{Bradlyn2016}, but not identified as such).
Since the RS fermion is nonchiral, it can, but need not, display surface Fermi arcs (the same is true for Dirac fermions\cite{Kargarian2016,Le2018}).

Our work is applicable to materials with commensurate magnetic order that breaks time-reversal symmetry but preserves the product of time-reversal and a spatial symmetry. We consider the limit where the magnetic order is ``frozen,'' such that the system is described by weakly interacting single-particle excitations. We leave the treatment of incommensurate magnetic ordering and dynamical magnetic moments to future work.

The paper is outlined as follows: in Sec.~\ref{sec:magnetic} we define the magnetic space groups.
In Sec.~\ref{sec:irreps} we derive the magnetic space groups that can host 3-, 6-, and 8-dimensional fermions.
In Sec.~\ref{sec:RS} we describe the Rarita-Schwinger fermion.
In Secs.~\ref{sec:kp3} and \ref{sec:kp6} we present the $\mathbf{k}\cdot\mathbf{p}$ Hamiltonians that describe the threefold and sixfold degenerate fermions.
We present candidate material realizations in Sec.~\ref{sec:materials}.
A discussion and outlook are in Sec.~\ref{sec:discussion}.


\section{Magnetic groups and corepresentations}
\label{sec:magnetic}

Of the 1421 magnetic space groups mentioned in Sec.~\ref{sec:intro}, 230 of them correspond to the ordinary space groups with no anti-unitary symmetry operations. 
Since these ``Type I" groups do not contain time-reversal symmetry, they can describe crystals with commensurate magnetic order.
However, they describe only a limited set of magnetic orderings.
A more complete set is given by space groups that lack time reversal symmetry, but do contain other anti-unitary elements.
For a thorough introduction to the magnetic space groups, we refer the reader to Refs.~\onlinecite{Cracknell,LitvinBook,MillerLove}.
Here, we give a brief overview of the different types of space groups, with and without anti-unitary symmetries, following the notation of Ref~\onlinecite{Cracknell}.

Given a Type I group, $H$, adding time-reversal symmetry, $\mathcal{T}$, as a generator yields a Type II (nonmagnetic) group, $H\cup \mathcal{T}H$.
Thus, there are also 230 Type II groups.

\begin{figure}[t]
\includegraphics[width=0.3\textwidth]{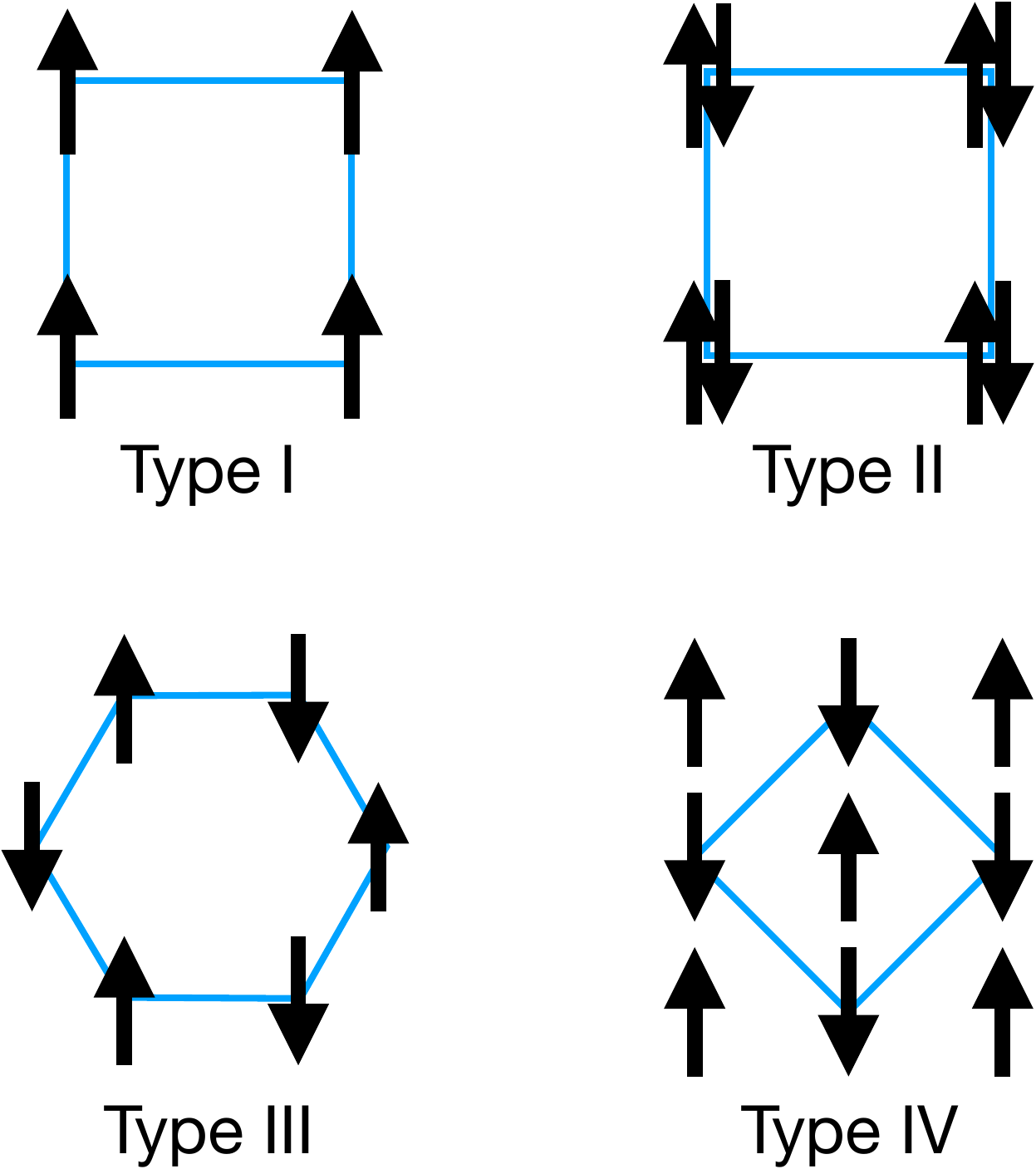}
\caption{Examples of Type I, II, III and IV groups. A Type I group does not contain any anti-unitary elements. A Type II group contains time-reversal and consequently every site must have both spin degrees of freedom. A Type III group contains the product of time-reversal and a unitary symmetry; in this example a $C_6$ rotation followed by time-reversal is a symmetry of the lattice. A Type IV group contains the product of time-reversal and a fractional lattice translation; in this example a translation of $\hat{\mathbf{x}}$ followed by time-reversal is a symmetry of the lattice, but $\hat{\mathbf{x}}$ is not itself a lattice vector. Type III and IV groups necessarily have more than one site in the unit cell. In all cases the blue square or hexagon outlines the unit cell.}
\label{fig:magnetictypes}
\end{figure}

The remainder of the magnetic space groups contain an anti-unitary generator that is the product of time-reversal and a unitary symmetry operation, but do not possess time-reversal symmetry itself.
These groups can be written in the form
\begin{equation}
G= H \cup \mathcal{T} g_0H,
\label{eq:defG}
\end{equation} 
where $H$ is a Type I group and $g_0\notin H$ is a unitary element.
If the set $g_0H$ does not (does) contain a fractional lattice translation, the group is referred to as a Type III (Type IV) group.
$H$ is referred to as a halving subgroup of $G$.
Examples of each Type are shown in Fig.~\ref{fig:magnetictypes}.

\begin{figure}





\includegraphics[width=0.3\textwidth]{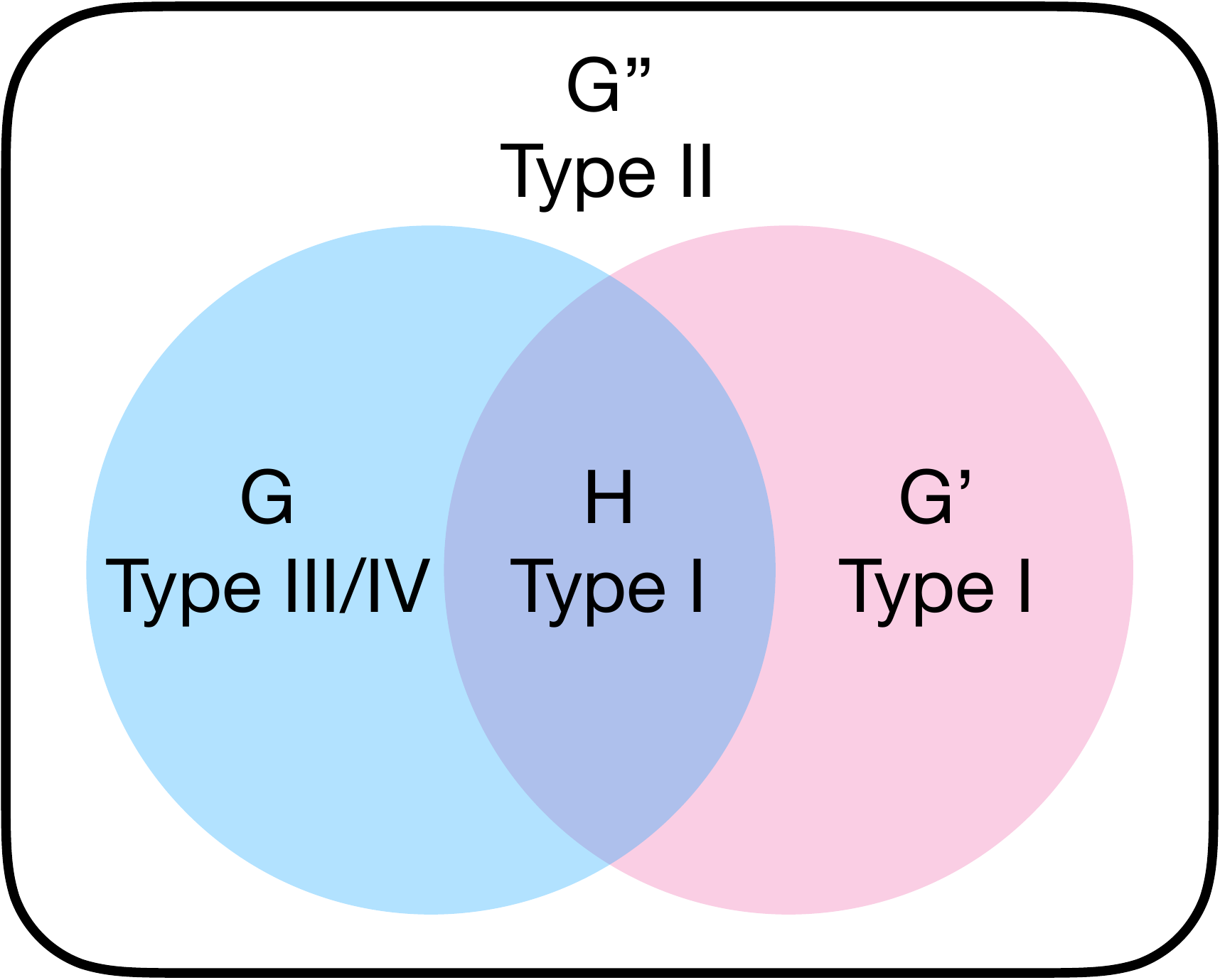}
\caption{Overlap of the groups $H, G, G'$ and $ G''$ defined in \cref{eq:defG,eq:defGpp,eq:defGp}. The blue circle indicates the Type III or IV magnetic group, $G$, while the pink circle indicates the Type I (unitary) group $G'$. Their intersection, $H=G\cap G'$, is exactly the unitary part of $G$.
The time-reversal-invariant Type II group $G''$ contains both $G$ and $G'$; the complement $G''\setminus (G\cup G')= \mathcal{T}H$ contains only antiunitary elements.}
\label{fig:groups}
\end{figure}

If we add time-reversal as a generator to $G$ we obtain the Type II group,
\begin{equation}
G'' = G\cup \mathcal{T}G = G'\cup \mathcal{T} G',
\label{eq:defGpp}
\end{equation}
where $G'$ is the Type I group defined by
\begin{equation}
G' = H \cup g_0 H.
\label{eq:defGp}
\end{equation}
$G''$ describes the symmetry of $G$ above the Neel temperature.
The relationship between $H, G, G'$ and $G''$ is depicted pictorially in Fig.~\ref{fig:groups}.

$G$ and $H$ contain the same lattice translations and hence have the same unit cell (even in the Type IV case, where $g_0H$ contains a fractional lattice translation, this translation by itself is not an element of $G$; it is only an element of $G$ when followed by $\mathcal{T}$.)
In the Type III groups, $G'$ and $G''$ have the same unit cell as $G$ and $H$, while in the Type IV groups, $G'$ and $G''$ have a different (smaller) unit cell than $G$ and $H$.

\subsection{Examples}
\label{sec:examples}

We present examples of $G, G', G''$ and $H$ in a Type III and Type IV group.

Let $G$ be the Type III group depicted in Fig.~\ref{fig:magnetictypes}.
$G$ is generated by the product of a six-fold rotation and time-reversal symmetry, as well as the translation vectors of the honeycomb lattice; the six-fold rotation, which is not itself a symmetry of $G$, plays the role of $g_0$ in Eq.~(\ref{eq:defG}). The unitary subgroup, $H$, of $G$, is generated by a three-fold rotation and translations.
In contrast, the unitary group $G'$ is generated by the six-fold rotation and translations (and therefore contains elements that are not in $G$, as depicted in the Venn diagram in Fig.~\ref{fig:groups}.)
The Type II supergroup $G''$ has the same generators as $G'$ and, in addition, time-reversal symmetry; therefore $G''$ contains $G$, $G'$ and $H$ as subgroups.

As a second example, let $G$ be the Type IV group depicted in Fig.~\ref{fig:magnetictypes}.
$G$ is generated by a four-fold rotation about the origin, translations by $\hat{\mathbf{x}} \pm \hat{\mathbf{y}}$, and the product of time-reversal and a translation by $\hat{\mathbf{x}}$ (the translation by $\hat{\mathbf{x}}$ plays the role of $g_0$ in Eq.~(\ref{eq:defG}) and is not a symmetry of $G$, since it moves an up spin to a down spin in Fig.~\ref{fig:magnetictypes}.)
The unitary subgroup, $H$, of $G$, is generated by a four-fold rotation and translations by $\hat{\mathbf{x}} \pm \hat{\mathbf{y}}$.
The unitary group $G'$ is generated by the generators of $H$ and, in addition, the translation by $\hat{\mathbf{x}}$. Hence, $G'$ has a smaller unit cell than $G$, as is always the case in a Type IV group.
The Type II supergroup $G''$ has the same generators as $G'$, in addition to time-reversal symmetry.


\subsection{Magnetic group labelling conventions}
\label{sec:notation}
There are two conventions to describe Type III and Type IV groups, the Belov-Nerenova-Smirnova\cite{BNS} (BNS) and the Opechowski-Guccione\cite{OG} (OG) notations.
We utilize the BNS notation because it more naturally accommodates the magnetic Brillouin zone, which is crucial for correctly determining degeneracies in the band structure of magnetic compounds.
While we again refer the reader to Refs.~\onlinecite{Cracknell,LitvinBook,MillerLove} for details, we briefly summarize the BNS notation here.
Each magnetic space group is assigned both a symbolic label (e.g. $P4_3'32'$ or $P_I2_13$) and a numeric label (e.g. 212.61 or 198.11).
The symbolic label uses the Hermann-Mauguin notation for space groups, with two additions: 1) a prime after an operation indicates that the operation is not in the space group, but the product of that operation and time-reversal symmetry is, and 2) in the Type IV groups, the lattice has a subscript indicating the magnetic Bravais lattice.
The numeric label consists of two numbers: in the Type III groups, the number before the decimal point corresponds to the unitary group $G'$, while in Type IV groups, the number before the decimal point corresponds to the unitary subgroup $H$.
The number after the decimal point is an index.


\section{Irreducible coreps of $G_\mathbf{k}$}
\label{sec:irreps}

We are interested in degeneracies that occur at high-symmetry points in the Brillouin zone. 
An $n$-fold degeneracy at a point $\mathbf{k}$ occurs if there is an $n$-dimensional irreducible corepresentation of the little group, $G_\mathbf{k}$, which consists of all symmetry operations that leave $\mathbf{k}$ invariant modulo a reciprocal lattice vector (an equivalence denoted by ``$\equiv$"): 
\begin{equation} 
G_\mathbf{k} \equiv \{ g \in G | g\mathbf{k} \equiv \mathbf{k} \},
\label{eq:littlegroup}
\end{equation}
where equality in momentum space is defined modulo a reciprocal lattice vector and translations act trivially in momentum space.
Corepresentations are the generalization of representations to groups with anti-unitary elements; for simplicity we will refer to irreducible corepresentations as irreps, whether or not they contain antiunitary elements.


In the Type I groups, $G_\mathbf{k}$ is always unitary and its irreps are easily looked up on the Bilbao Crystallographic Server (BCS)\cite{GroupTheoryPaper} or in Ref.~\onlinecite{Cracknell}.
Previously\cite{Bradlyn2016}, we used these tables to find the high-dimensional irreps for the Type I and II groups; the results are reproduced in Table~\ref{tab:NF}.
However, the analogous searchable tables do not yet exist for magnetic groups.
In this section, we explain how to efficiently find the irreducible coreps of dimension 3-, 6- and 8- in the Type III and IV groups from the tables in Ref.~\onlinecite{MillerLove}, which list the dimensionality of all irreps at all high-symmetry points in the remaining 1191 magnetic space groups.
We need only consider the high-symmetry points because their little groups contain the little groups of all other points in the BZ.
We focus exclusively on the ``double-valued'' irreps, where $2\pi$ rotations are represented by $-\mathbb{I}$, ($\mathbb{I}$ denotes the identity matrix); these describe systems with non-negligible spin-orbit coupling, which we expect to apply to many magnetic systems.

\begin{table}[ht]
    \centering
    \begin{tabular}{ccc||ccc}
         Deg. & SG & $\mathbf{k}$ & Deg. & SG & $\mathbf{k}$\\
         \hline
         3 & $I2_13$ (199) & P & 8 & $P4/ncc$ (130) & A\\
         3 & $I4_132$ (214) & P & 8 & $P4_2/mbc$ (135) & A\\
         3 & $I\bar{4}3d$ (220) & P & 8 & $P\bar{4}3n$ (218) & R\\
         6 & $P2_13$ (198) & R & 8 & $I\bar{4}3d$ (220) & H\\
         6 & $Pa\bar{3}$ (205) & R & 8 & $Pn\bar{3}n$ (222) & R\\
         6 & $Ia\bar{3}$ (206) & P & 8 & $Pm\bar{3}n$ (223) & R\\
         6 & $P4_332$ (212) & R & 8 & $Ia\bar{3}d$ (230) & H\\
         6 & $P4_132$ (213) & R & \\
         6 & $Ia\bar{3}d$ (230) & P & 
    \end{tabular}
    \caption{Type II space groups that display 3-, 6- and 8-fold fermions with spin-orbit coupling, reproduced from Ref.~\onlinecite{Bradlyn2016}. The first column indicates the degeneracy, the second column the space group, and the third the high-symmetry point. The same columns are repeated to the right of the double-line for the 8-fold degeneracies.
    In the Type I groups, the 3-fold crossings remain 3-fold, the 6-fold crossings split into two 3-folds, and the 8-fold crossings split into two 4-folds.}
    \label{tab:NF}
\end{table}

\subsection{Type III groups}
In the Type III groups, $G, H, G'$ and $G''$ share the same BZ.
Thus, if there is a 3-, 6-, or 8-fold degeneracy at $\mathbf{k}$ in $G$, then there must also be a 3-, 6-, or 8-fold degeneracy at $\mathbf{k}$ in the Type II group $G''$, which has more symmetry than $G$.
The Type II groups with 3-, 6- and 8-fold degeneracies were enumerated by us in Ref.~\onlinecite{Bradlyn2016}, and reproduced in Table~\ref{tab:NF} for convenience.
Consequently, we can find all magnetic Type III space groups with these degeneracies by scanning the tables in Ref.~\onlinecite{MillerLove} for Type III groups whose numeric label starts with one of the $G''$ listed in Table~\ref{tab:NF} (recall that $G''$ and $G'$ have the same numeric label and that the tables in Ref.~\onlinecite{MillerLove} are sorted in BNS notation, which lists Type III groups by $G'$.)
The results are shown in Table~\ref{table:magneticNF}.

\subsection{Type IV groups}
We need to be more clever for the Type IV groups because their BNS label contains $H$, but not $G'$.
In addition, since $G'$ and $G$ have different Brillouin zones, it is not immediately evident how to find the irreps of $G_\mathbf{k}$ from those of $G'_\mathbf{k}$.

We start by relating irreps of $G_\mathbf{k}$ to those of $H_\mathbf{k}$.
As discussed in Sec.~\ref{sec:magnetic}, $G$ and $H$ always have the same BZ.
Hence, $H_\mathbf{k}$ is exactly the unitary part of $G_\mathbf{k}$ for every $\mathbf{k}$.
Thus, if $G_\mathbf{k}$ is unitary, then $G_\mathbf{k} = H_\mathbf{k}$. It follows that the irreps of $G_\mathbf{k}$ (and their dimensionality) are exactly the irreps of $H_\mathbf{k}$, which can be looked up on the BCS.

Now suppose $G_\mathbf{k}$ contains anti-unitary elements. Then there must exist an $h_0\in H$ such that 
\begin{equation}
\mathcal{T} g_0 h_0 \mathbf{k} \equiv \mathbf{k},
\label{eq:defh0}
\end{equation}
where $g_0$ is defined in Eq.~(\ref{eq:defG});
otherwise, $G_\mathbf{k}$ would be unitary.
We prove in Appendix~\ref{sec:prooflittlegroup} that $G_\mathbf{k}$ takes the form
\begin{equation}
G_\mathbf{k} = H_\mathbf{k} \cup \mathcal{T} g_0 h_0 H_\mathbf{k}.
\label{eq:TypeIIIanti}
\end{equation}
It follows that the dimensionality of the irreps of $G_\mathbf{k}$ are determined by examining each irrep of $H_\mathbf{k}$ and determining whether its dimension stays the same or doubles when an anti-unitary generator ($\mathcal{T}g_0h_0$) is added to the group.
The algorithm for determining this is detailed in Refs.~\onlinecite{Cracknell} and \onlinecite{MillerLove}.
However, because implementing the algorithm directly is time consuming, we now explain how to use the existing tables\cite{MillerLove} to efficiently search the Type IV magnetic groups for a group, $G$, and high-symmetry point, $\mathbf{k}$, such that $G_\mathbf{k}$ has a 3-, 6-, or 8-dimensional irrep.

Since each irrep of $G_\mathbf{k}$ either has the same or double the dimensionality of an irrep of $H_\mathbf{k}$, a necessary condition for $G_\mathbf{k}$ to have a 3-, 6-, or 8-dimensional irrep is that $H_\mathbf{k}$ have a 3- or 4-dimensional irrep
(double-valued irreps of $H_\mathbf{k}$ are at most 4-dimensional.\cite{Cracknell,Bradlyn2016,GroupTheoryPaper})
We have enumerated the $H_\mathbf{k}$ with 3- or 4-dimensional double-valued irreps in Tables~\ref{table:3ddouble} and \ref{table:4ddouble} in Appendix~\ref{sec:listdoubleirreps}.
For reference, we have also enumerated the 3-, 4-, and 6-dimensional single-valued irreps in Appendix~\ref{sec:listsingleirreps}.
These can be used to extend our results to magnetic materials with negligible spin-orbit coupling.

The tables in Ref.~\onlinecite{MillerLove} list the irreps of the magnetic space groups at all high-symmetry points in BNS notation, which, for Type IV groups, is sorted by the unitary subgroup $H$.
Thus, for each $H$ and $\mathbf{k}$ in Table~\ref{table:3ddouble} or \ref{table:4ddouble}, we search Ref.~\onlinecite{MillerLove} for the magnetic groups derived from $H$ and determine which of these have the 3-, 6-, or 8-dimensional double-valued irreps we seek.
The results are in Table~\ref{table:magneticNF}.

\subsection{Summary of results} 
The cases where $G_\mathbf{k}$ has a 3-, 6-, or 8-dimensional irrep or corep are listed in Table~\ref{table:magneticNF}. 

In the following sections, we describe the results in more detail. In particular, in Sec.~\ref{sec:RS} we prove which of the 8-dimensional irreps describe Rarita-Schwinger fermions.
In Sec.~\ref{sec:kp3}, we prove that all of the 3-dimensional irreps carry a charge-2 monopole of Berry curvature, which we named a ``spin-1 Weyl'' in Ref.~\onlinecite{Bradlyn2016}; consequently, they are accompanied by surface Fermi arcs. 
(All magnetic threefold degeneracies are non-degenerate away from the nodal point, in contrast to some nonmagnetic cases.)
In Sec.~\ref{sec:kp6}, we show that all of the 6-fold fermions are nonchiral and topologically equivalent to two opposite-chirality copies of the spin-1 Weyl; we appropriately refer to them as ``spin-1 Dirac'' fermions.
{In general, we find that the Type IV groups have an extra symmetry constraint that pins the $\mathbf{k} \cdot \mathbf{p}$ Hamiltonian to an exactly solvable point, which would not be achievable in the non-magnetic space groups without fine-tuning.}

\begin{table*}

\subfloat[3-fold degeneracies]{
\begin{tabular}{cccccc}
Type & $G$ & BNS & $H$ & $G'$ & $\mathbf{k}$\\
\hline
III & $P4_3'32'$ & 212.61 & 198 & 212 & $R$\\
III & $P4_1'32'$ & 213.65 & 198 & 213 & $R$\\
III & $I4_1'32'$ & 214.69 & 199 & 214 & $P$\\
III & $I\bar{4}'3d'$ & 220.91 & 199 & 220 & $P$\\
III & $Ia\bar{3}d'$ & 230.148 & 206 & 230 & $P$\\
\hline
IV & $P_I2_13$ & 198.11 & 198 & 199 & $R$\\
IV & $P_I4_3 32$ & 212.62 & 212 & 214 & $R$\\
IV & $P_I4_1 32$ & 213.66 & 213 & 214 & $R$
\end{tabular}
}
\quad\quad
\subfloat[6-fold degeneracies]{
\begin{tabular}{cccccc}
Type & $G$ & BNS & $H$ & $G'$ & $\mathbf{k}$\\
\hline
III & $Pa'\bar{3}'$ & 205.35 & 198 & 205 & $R$\\
III & $Ia'\bar{3}'$ & 206.39 & 199 & 206 & $P$\\
III & $Ia'\bar{3}'d$ & 230.147 & 220 & 230 & $P$\\
III & $Ia'\bar{3}'d'$ & 230.149 & 214 & 230 & $P$\\
\hline
IV & $P_Ia\bar{3}$ & 205.36 & 205 & 206 & $R$
\end{tabular}
}
\quad\quad
\subfloat[8-fold degeneracies]{
\begin{tabular}{cccccc}
Type & $G$ & BNS & $H$ & $G'$ & $\mathbf{k}$\\
\hline
III & $Pn'\bar{3}'n'$ & 222.102 & 207 & 222 & $R$\\
III & $Pm'\bar{3}'n'$ & 223.108 & 208 & 223 & $R$\\
III & $Ia'\bar{3}'d'$ & 230.149 & 214 & 230 & $H$\\
\hline
IV & $P_I4/nbm$ & 125.374 & 125 & 140 & $A$\\ 
IV & $P_C4/nnc$ & 126.385 & 126 & 124 & $A$\\
IV & $P_C4/nmm$ & 129.420 & 129 & 129 & $A$\\
IV & $P_C4_2/mmc$ & 131.445 & 131 & 132 & $A$\\
IV & $P_I4_2/mcm$ & 132.458 & 132 & 140 & $A$\\
IV & $P_C4_2/mnm$ & 136.504 & 136 & 127 & $A$\\
IV & $P_I\bar{4}3m$ & 215.73 & 215 & 217 & $R$\\
IV & $P_Im\bar{3}m$ & 221.97 & 221 & 229 & $R$\\
IV & $P_In\bar{3}m$ & 224.115 & 224 & 229 & $R$\\
\end{tabular}
}
\caption{Magnetic space groups, $G$, with 3-, 6- and 8-dimensional irreps.
The first column in each table indicates whether the group is Type III or Type IV; the second column gives the symbolic name of the group in BNS notation; the third column gives the space group number in BNS notation; the fourth column gives the number of the unitary subgroup; the fifth column gives the number of the corresponding Type I group, $G'$, (Eq.~(\ref{eq:defGp})); and the sixth column gives the $\mathbf{k}$ point where the 3-dimensional irrep is located.
Each group would be listed in the ICSD under $G'$, which describes the unitary part of the enlarged space group when time-reversal symmmetry is restored.}
\label{table:magneticNF}
\end{table*}

\section{Rarita-Schwinger fermions}
\label{sec:RS}
A generalization of Dirac fermions, Rarita-Schwinger (RS) fermions are relativistic particles transforming in the spin-$3/2$ representation of the rotation group\cite{raritaschwinger}. Although never observed as fundamental particles, RS fermions play an essential role in supergravity: the supersymmetric partner to the graviton is predicted to be a RS particle\cite{sugra}. The Hamiltonian of a massless RS fermion takes the little-known form\cite{moldauer-hamiltonian}
\begin{equation}
H_{RS}=\tau_z\otimes (\mathbf{k}\cdot\mathbf{J})\left(\frac{13}{4}-\frac{(\mathbf{k}\cdot\mathbf{J})^2}{|\mathbf{k}|^2}\right), \label{eq:rs}
\end{equation}
where $\mathbf{J}$ is the vector of spin-$3/2$ matrices, and $\tau_z$ is a Pauli matrix acting in the chirality basis. Note that since the spin-$3/2$ matrices do not all anticommute, the second term in the parentheses is \emph{non-analytic} in $\mathbf{k}$. Because of this, we should not expect to reproduce this term in a $\mathbf{k}\cdot\mathbf{p}$ expansion for a crystal Hamiltonian. It has recently become fashionable\cite{Tang2017,Liang2016-1,Flicker2018} to refer to quasiparticles described by the Hamiltonian
\begin{equation}
H_{RS*}=\tau_z\otimes(\mathbf{k}\cdot\mathbf{J}),\label{eq:rsstar}
\end{equation}
which corresponds to two decoupled spin-$3/2$ fermions of opposite chirality, as RS fermions; we shall refer to these as RS* fermions to emphasize the differences. 
While the Hamiltonians $H_{RS}$ and $H_{RS*}$ share the same eigenstates, they are not topologically equivalent. 
To see this, let us focus on the $+$ chirality. Using the results of Refs.~\onlinecite{Berry1984,Bradlyn2016}, we see that the positive energy bands in the RS Hamiltonian are degenerate (as mandated by Lorentz invariance), with Chern numbers $+3$ and $+1$ (we define the Chern number of a band to be the Chern number of a putative Fermi surface which resides in that band). 
In the RS* Hamiltonian, on the other hand, all bands are non-degenerate away from $\mathbf{k}=0$. Though distinct, both of these are homotopic to different phases of the ``spin-$3/2$ fermion'' introduced in Ref.~\onlinecite{Bradlyn2016}. 
In Ref.~\onlinecite{Bradlyn2016}, the RS spectrum was realized as the critical point in the phase diagram of the spin-3/2 fermion, while the RS* Hamiltonian was realized at a fine-tuned point.

We will now search for quasiparticles near band degeneracies in crystals that can reproduce the dynamics of the $8\times 8$ RS and RS* Hamiltonians.
Let us first revisit the $\mathbf{k}\cdot\mathbf{p}$ Hamiltonians for the eightfold degeneracies in the \emph{nonmagnetic} Type II space groups $Pn\bar{3}n1'$ (222) and $Pm\bar{3}n1'$ (223). While we will primarily reproduce the results of Ref.~\onlinecite{Bradlyn2016}, we will do so in a way which generalizes to a treatment of multifold degeneracies in other (magnetic) space groups, and sheds new light on some of the complicated $\mathbf{k}\cdot\mathbf{p}$ Hamiltonians presented there. 

To begin, we note that the two Type II space groups $\sgtwelve$ and $\sgthirteen$ (the $1'$ indicates that time-reversal symmetry is a generator) can be expressed in terms of Type II halving subgroups,
\begin{align}
\sgtwelve&\approx \sgseven\cup\left\{I|\half\half\half\right\}\sgseven\label{eq:sg222decomp} \\
\sgthirteen&\approx \sgeight\cup\left\{I|0\right\}\sgeight \label{eq:sg223decomp},
\end{align}
where ``$\approx$'' denotes a group isomorphism and we have used the standard notation\cite{Cracknell} where $\{ R|\mathbf{t}\}$ denotes a point group operation $R$ following by a translation $\mathbf{t}$; $I$ indicates the inversion symmetry operation. 
In Table~\ref{table:2223degens}, we list the high symmetry points in these space groups with 3-, 4-, 6- and 8-fold degeneracies, and their degeneracy with and without spin-orbit coupling.
\begin{table}[t]
\centering
\begin{tabular}{c|c|c}
$\mathbf{k}$ & No SOC & SOC\\
\hline
$\Gamma$ & $3$ & $4$ \\
$R$ & $6$ & $8$ \\
$M$ & - & $4$ \\
$X$ & $4$ & $4$
\end{tabular}
\caption{3-, 4-, 6- and 8-fold degeneracies in the nonmagnetic Type II space groups $\sgtwelve$ and $\sgthirteen$.}
\label{table:2223degens}
\end{table}
Let us focus first on the group $\sgtwelve$ (SG 222); the halving subgroup in this case is, from Eq.~(\ref{eq:sg222decomp}), the symmorphic space group $\sgseven$ (SG 207), with point group $O$. Because this group is symmorphic, we can determine its little group representations at any $\mathbf{k}$ point by examining the commutation relations between the generators of the point group $O=\langle C_{4z}, C_{3,111}\rangle$, $\mathcal{T}$, and the coset representative $g=\left\{I|\half\half\half\right\}$. In particular, 
\begin{align}
gC_{4z}g^{-1}&=\{C_{4z}|100\}\label{eq:222coms1} \\
gC_{3,111}g^{-1}&=C_{3,111} \label{eq:222coms3} \\
g\mathcal{T}g^{-1}&=\mathcal{T}\label{eq:222coms2}
\end{align}
For any point $\mathbf{k}$ which is not a TRIM, the element $g$ will not be in the little group $G_\mathbf{k}$. Thus, for these $\mathbf{k}$ points, the little groups in space groups $\sgseven$ and $\sgtwelve$ coincide. 
For the TRIM points, however, the size of the little group in SG $\sgtwelve$ is doubled relative to SG $\sgseven$.  \cref{eq:222coms1,eq:222coms3,eq:222coms2} show 
that at a TRIM point, $\mathbf{k}$, a representation, $\dk$, 
of the little group in SG $\sgtwelve$ satisfies the following commutation relations;
\begin{align}
\dk(g)\dk(\mathbf{t}_i)\dk(g)^{-1}&=\dk(-\mathbf{t}_i)\label{eq:comm1} \\
\dk(g)\dk(C_{4z})\dk(g)^{-1}&=e^{-i\mathbf{k}\cdot\mathbf{t}_1}\dk(C_{4z})\label{eq:comm2} \\
\dk(g)\dk(C_{4x})\dk(g)^{-1}&=e^{-i\mathbf{k}\cdot\mathbf{t}_2}\dk(C_{4x}) \label{eq:comm3}\\
\dk(g)\dk(C_{4y})\dk(g)^{-1}&=e^{-i\mathbf{k}\cdot\mathbf{t}_3}\dk(C_{4y}) \label{eq:comm4}\\
\dk(g)\dk(C_{3,111})\dk(g)^{-1}&=\dk(C_{3,111}) \label{eq:comm5}\\
\dk(g)\dk(\mathcal{T})\mathcal{K}\dk(g)^{-1}&=\dk(\mathcal{T})\mathcal{K},\label{eq:comm6}
\end{align}
where $\mathcal{K}$ is the complex conjugation operation, and $\{\mathbf{t}_i\}$ are a basis for the cubic Bravais lattice. Since we are primarily interested in the eightfold degenerate fermion in this space group, let us focus on the $R$ point. 
To prove that the low-energy theory near $R$ is described by the $RS$ or $RS*$ Hamiltonian, 
we will build the eight-dimensional representation $\Delta_\mathbf{k}$ out of the four-dimensional spin-3/2 representation of the $R$ point in $\sgseven$.
We will show how the inclusion of $g$ in the little group requires the degeneracy of the spin-3/2 representation to double, and that the $\mathbf{k}\cdot\mathbf{p}$ Hamiltonian takes a RS form.

\subsection{Doubling of the Degeneracy} 
\label{sec:doubledeg}

The reduced coordinates of the $R$ point are $(\half,\half,\half)$. 
The little group at $R$, which we denote by $G^{222}_R$ (the superscript $222$ indicates the numeric symbol for $\sgtwelve$), contains the entirety of space group $\sgseven$, as well as the coset representative $g$. 
By taking $\mathbf{k} = (\half,\half,\half)$ in \cref{eq:comm1,eq:comm2,eq:comm3,eq:comm4,eq:comm5,eq:comm6}, the commutation relations
show that at the $R$ point, the representative of $g$ must commute with the representative of $C_{3,111}$, and anticommute with the representatives of $C_{4x},C_{4y}$ and $C_{4z}$. Let us focus on the spin-$3/2$ representation, $\rho$, of the little group $G_R^{207}$ (the superscript $207$ indicates the numerical symbol for $\sgseven$), and use it to build a representation $\Delta$ of $G_R^{222}$. 
We will show first that we cannot consistently define a four-dimensional representation matrix for $g$, and hence conclude that the fourfold degeneracy at $R$ in space group $\sgseven$ doubles when going to space group $\sgtwelve$. For convenience, we introduce the spin-$3/2$ matrices,
\begin{align}
J_x&=\frac{1}{2}\left(
\begin{array}{cccc}
0 & \sqrt{3} & 0 & 0 \\
\sqrt{3} & 0 & 2 & 0 \\
0 & 2  & 0 & \sqrt{3} \\
0 & 0 & \sqrt{3} & 0
\end{array}
\right)\\
J_y&=-\frac{i}{2}\left(
\begin{array}{cccc}
0 & \sqrt{3} & 0 & 0 \\
-\sqrt{3} & 0 & 2 & 0 \\
0 & -2 & 0 & \sqrt{3} \\
0 & 0 & -\sqrt{3} & 0
\end{array}
\right)\\
J_z&=\frac{1}{2}\left(
\begin{array}{cccc}
3 & 0 & 0 & 0 \\
0 & 1 & 0 & 0 \\
0 & 0 & -1 & 0 \\
0 & 0 & 0 & -3
\end{array}
\right)
\end{align}

We also introduce two-by-two Pauli matrices $\sigma_i$ and $\tau_i$, such that the $\tau_i$ exchange the two-by-two blocks in the spin-3/2 basis, and the $\sigma_i$ act within the blocks. Using these matrices, the spin-$3/2$ representation $\rho$ can be written as,
\begin{align}
\rho(C_{3,111})&=\exp\left[-\frac{2\pi i}{3\sqrt{3}}\left(J_x+J_y+J_z\right)\right] \\
\rho(C_{4z})&=\exp\left(-\frac{i\pi}2J_z\right)=-\frac{1}{\sqrt{2}}\left(\sigma_z\tau_z+i\sigma_0\tau_z\right) \\
\rho(\mathcal{T})&=\exp\left(-i\pi J_y\right)\mathcal{K}=\sigma_y\tau_x\mathcal{K}
\end{align}
Since $g^2=E$, and since $\rho(g)$ must anticommute with $\rho(C_{4z})$, it must be that
\begin{equation}
\rho(g)\stackrel{?}{=}(\alpha\sigma_z+\beta\sigma_0)\otimes(\gamma\tau_x+\delta\tau_y)
\end{equation}
Since $\rho(g)$ must also commute with $\rho(\mathcal{T})$, it must be that $\alpha =0$, and so our putative $\rho(g)$ takes the form
\begin{equation}
\rho(g)\stackrel{?}{=}\gamma\tau_x+\delta\tau_y.
\end{equation}
However, upon attempting to enforce the constraint
\begin{equation}
[\rho(g),\rho(C_{3,111})]=0,
\end{equation}
we find $\delta=\gamma=0$. Thus, we conclude that there does not exist a four-dimensional representation of $G^{222}_R$ in space group $\sgtwelve$ that reduces to the spin-3/2 representation in space group $\sgseven$. Therefore, upon adding the symmetry $g$ to space group $\sgseven$, the four dimensional representation at $R$ must double in size. One convenient choice of basis for this eight-dimensional irreducible representation $\Delta$ in space group $\sgtwelve$ is given by
\begin{align}
\Delta(C_{4z})&=\exp\left(-\frac{i\pi}2J_z\right)\otimes\mu_z \\
\Delta(C_{3,111})&=\exp\left[-\frac{2\pi i}{3\sqrt{3}}\left(J_x+J_y+J_z\right)\right]\otimes\mu_0 \\
\Delta(\mathcal{T})&=\exp\left(-i\pi J_y\right)\otimes\mu_0\mathcal{K} \\
\Delta(g)&=\mu_x,
\end{align}
where we have introduced a new set of Pauli matrices $\mu$ which act to exchange the different copies of $\rho$ in this eight-dimensional irrep. It is clear that by breaking inversion symmetry and reducing $G^{222}_R$ to $G_R^{207}$, the representation $\Delta$ restricts to
\begin{equation}
\Delta\downarrow G_R^{207}=\rho\oplus\rho\label{eq:222to207}
\end{equation}

\subsection{RS Hamiltonian}
\label{sec:RSHam}

We have thus constructed an eight-dimensional representation at the $R$ point in space group $\sgtwelve$ in a basis that makes explicit the important of the inversion symmetry $g$. We will now impose these symmetry constraints on the $\mathbf{k}\cdot\mathbf{p}$ Hamiltonian expanded about $R$. We will see that our choice of basis makes the connection to the RS and RS* Hamiltonian manifest.

To begin, let us write the $\mathbf{k}\cdot \mathbf{p}$ Hamiltonian near the $R$ point as
\begin{equation}
H_R(\mathbf{k})=\left(\begin{array}{cc}
A(\mathbf{k}) & B(\mathbf{k}) \\
B^\dag(\mathbf{k}) & C(\mathbf{k})
\end{array}
\right),
\end{equation}
in terms of the $4\times 4$ block matrices $A,B$ and $C$. Next, From Eq.~(\ref{eq:222to207}) we deduce that
\begin{equation}
A(\mathbf{k})=H_{207}(\mathbf{k}),
\end{equation}
where $H_{207}(\mathbf{k})$ is the spin-3/2 generalized $\mathbf{k}\cdot \mathbf{S}$ Hamiltonian presented in Ref.~\onlinecite{Bradlyn2016}. It can be in either the RS or RS* phase. The next simplest symmetry to impose is the composite $g\mathcal{T}$, which yields the constraint
\begin{equation}
H_R(\mathbf{k})=\sigma_y\tau_x\mu_xH^*_R(\mathbf{k})\sigma_y\tau_x\mu_x,\label{eq:H222r}
\end{equation}
which implies that
\begin{align}
C(\mathbf{k})&=\sigma_y\tau_xH_{207}^{*}(\mathbf{k})\sigma_y\tau_x\label{eq:222C} \\
B^T(\mathbf{k})&=\sigma_y\tau_x B(\mathbf{k})\sigma_y\tau_x\label{eq:222B}
\end{align}
Eq.~(\ref{eq:222C}) fully determines $C(\mathbf{k})$: we see that if $B(\mathbf{k})=0$, then the Hamiltonian $H_R(\mathbf{k})$ would be homotopic to the nonchiral RS and RS* Hamiltonians. However, Eq.~(\ref{eq:222B}), combined with the constraints of $C_3$ symmetry, also allow for a single nontrivial solution:
\begin{equation}
B(\mathbf{k})=c\left(k_x\sigma_y\tau_z+k_y\sigma_x\tau_z+k_z\tau_y\right)\mu_y.\label{eq:222couple}
\end{equation}

Thus, we see that the general $\mathbf{k}\cdot\mathbf{p}$ Hamiltonian for the eightfold degeneracy at the $R$ point of space group $\sgtwelve$ consists of two spin-3/2 fermions of opposite chirality, coupled with the $\mathbf{k}$-dependent matrix $B(\mathbf{k})$. In the language of high-energy physics, this corresponds to a nonchiral RS or RS* fermion with a $\mathbf{k}$-dependent mass $B(\mathbf{k})\mu_y$. The matrix $\mu_z$ is the chirality matrix, which, due to the presence of the mass term, is not a conserved quantity, unless $c=0$. When $c$ is not zero, the chirality of a given state is a time-dependent quantity which oscillates with momentum-dependent frequency $\omega_C=2c|\mathbf{k}|$.

The essential ingredients for the preceding analysis in this section to hold are a cubic point group and the combined antiunitary symmetry $g\mathcal{T}$. Consequently, if $g$ and $\mathcal{T}$ are broken by magnetism, but their product is preserved, the eightfold degeneracy at $R$ need not split. Instead, the form of the coupling $B(\mathbf{k})$ will change.

\subsection{Space groups that can host nonchiral RS fermions}
To fully enumerate the RS fermions that can occur in crystals, we now examine the eightfold degeneracies in other magnetic and nonmagnetic space groups. We begin with the nonmagnetic groups.

\subsubsection{Nonmagnetic}

We showed in Ref.~\onlinecite{Bradlyn2016} {(concurrently with Ref.~\onlinecite{Wieder2016})} that there are seven nonmagnetic Type II space groups with eightfold degeneracies in the presence of SOC; these are listed in Table~\ref{table:8ddouble}. We now show that all of the eightfold fermions in the cubic groups are of the Rarita-Schwinger type, while those in the tetragonal space groups are distinct.

\begin{table}[t]
    \centering
    \begin{tabular}{cc}
        SG & $\mathbf{k}$ \\
        $P4/ncc1'$ (130) & $A$\\
        $P4_2/mbc1'$ (135) & $A$\\
        $P\bar{4}3n1'$ (218) & $R$ \\
        $I\bar{4}3d1'$ (220) & $H$ \\
        $Pn\bar{3}n1'$ (222) & $R$\\
        $Pm\bar{3}n1'$ (223) & $R$ \\
        $Ia\bar{3}d1'$ (230) & $H$ 
    \end{tabular}
    \caption{Non-magnetic space groups with time-reversal symmetry and SOC that display eightfold degeneracies, as shown in Refs.~\onlinecite{Bradlyn2016} and \onlinecite{Wieder2016}.}
    \label{table:8ddouble}
\end{table}

First, the little group of space group $\sgthirteen$ at the $R$ point is isomorphic to $G_R^{222}$. As such the analysis and $\mathbf{k}\cdot\mathbf{p}$ Hamiltonian derived in Secs.~\ref{sec:doubledeg} and \ref{sec:RSHam} also apply to the RS fermions in this space group. Next, recall that the Type II space groups $I\bar{4}3d1'$ (220) and $P\bar{4}3n1'$ (218) also host eightfold degeneracies, at the $H$ and $R$ points respectively; in Ref.~\onlinecite{Bradlyn2016} these were found to have a quite complicated $\mathbf{k}\cdot\mathbf{p}$ Hamiltonian. 
However, they yield the coset decompositions
\begin{align}
P\bar{4}3n1'&= P231' \cup \{m_{1\bar{1}0}|\half\half\half\} P231' \\
I\bar{4}3d1'&= I2_131' \cup \{m_{1\bar{1}0}|\frac{1}{4}\frac{3}{4}\frac{3}{4}\} I2_131',
\end{align}
which show that in both cases the eightfold fermions emerge from {an anti-chiral} doubling of the four dimensional irreducible corepresentation ${}^1\bar{F}{}^2\bar{F}$
of the structurally chiral tetrahedral group $T$ (isomorphic to the little cogroup of $P231'$ at the $R$ point, and $I2_13'$ at the $H$ point.)
As this corepresentation is the restriction of the spin-$3/2$ representation of the structurally chiral octahedral group $O$, we can follow the same logic as in Sec.~\ref{sec:RSHam} to find the $\mathbf{k}\cdot\mathbf{p}$ Hamiltonian
\begin{equation}
H_{218}(\mathbf{k})=\left(\begin{array}{cc}
H_{195}(\mathbf{k}) & B(\mathbf{k}) \\
B^\dag(\mathbf{k}) & \sigma_y\tau_xH_{195}^*(\mathbf{k})\sigma_y\tau_x
\end{array}
\right).
\end{equation}
Here $H_{195}(\mathbf{k})$ is the $\mathbf{k}\cdot \mathbf{p}$ Hamiltonian for the spin-3/2 fermions in tetrahedral space groups, first given in a different basis in Refs.~\onlinecite{Flicker2018,Chang2017}. It is a generalization of the Hamiltonian $H_{207}(\mathbf{k})$, with one additional free parameter, {due to the fact that SG 195 lacks $\{C_{4z} | 000\}$ as a generator}. 
We find that the chiral coupling $B(\mathbf{k})$ is a three-parameter generalization of Eq.~(\ref{eq:222couple}), whose explicit form is not particularly enlightening. Because the $\mathbf{k}\cdot\mathbf{p}$ Hamiltonian $H_{218}(\mathbf{k})$ reduces to $H_{R}(\mathbf{k})$ of Eq.~(\ref{eq:H222r}) for a particular choice of parameters, we conclude that space groups $P\bar{4}3n1'$ and $I\bar{4}3d1'$ can also host RS fermions. 

Finally, space group $\sgthirty$ also hosts an eightfold degeneracy. This space group is obtained by adding inversion symmetry to space group $I\bar{4}3d1'$. This restricts the $\mathbf{k}\cdot\mathbf{p}$ Hamiltonian in $\sgthirty$ to the form of Eq.~(\ref{eq:H222r}). Hence this space group hosts RS fermions as well.

Thus, we have shown that all eightfold fermions in the cubic, nonmagnetic space groups are of Rarita-Schwinger type. We now prove that the eightfold fermions in the tetragonal space groups $P4/ncc1'$ (130) and $P4_2/mbc1'$ (135) are fundamentally different. These space groups can also be written as a coset decomposition in terms of structurally chiral halving subgroups, which take the form
\begin{align}
P4/ncc1'&=P42_121'\cup \{I|000\} P42_12'1 \\
P4_2/mbc1' &= P4_22_121' \cup \{I|000\} P4_22_121'.
\end{align}
Each of these halving subgroups hosts a fourfold degenerate ``doubled spin-1/2'' fermion \cite{Flicker2018,kramersweyl,bouhon}, which is phenomenologically distinct from a spin-3/2 fourfold degeneracy (i.e. there are no Chern number $3$ bands, and no cubic rotational symmetries). 
Upon the addition of inversion symmetry, these double spin-1/2 degeneracies double again, yielding an eightfold degeneracy which was aptly named a double-Dirac fermion\cite{Wieder2016}. Thus, the tetragonal double Dirac fermions are distinct from Rarita-Schwinger fermions.

\subsubsection{Magnetic}
We proved in the previous section that every cubic Type II space group with eightfold degeneracies hosted Rarita-Schwinger fermions. 
The same is true for the Type III magnetic groups.
To see this, note that every cubic Type III magnetic space group with an eightfold degeneracy is a subgroup of a Type II space group which also has an eightfold degeneracy. One can thus obtain the $\mathbf{k}\cdot\mathbf{p}$ Hamiltonian for each Type III group by starting from the RS Hamiltonian in the Type II group, and relaxing symmetry constraints. This means that if the RS Hamiltonian is accessible as a point in parameter space for the Type II group, it must also be for the Type III group.

To conclude, we analyze the Type IV magnetic groups which host eightfold degeneracies. We will show below that these do not give RS fermions, but are still rather novel. There are three cases given in Table~\ref{table:magneticNF}. Each has a coset decomposition in terms of a Type IV structurally chiral halving subgroup:
\begin{align}
P_I\bar{4}3m&=P_I23\cup \{S_4|000\}P_I23 \\
P_Im\bar{3}m&=P_I432 \cup \{I|000\}P_I432 \\
P_In\bar{3}m&=P_I4_232 \cup \{I|000\}P_I4_232
\end{align}
In all three cases, the structurally chiral halving subgroup features a fourfold chiral fermion at the R point. Upon the addition of the orientation reversing symmetry (inversion or rotoinversion), the degeneracy doubles yielding a nonchiral eightfold degenerate fermion. We illustrate this for only the simplest case, $P_Im\bar{3}m$, as all three cases are similar. At the $R$ point, the little group of the halving subgroup $P_I432$ has the following four dimensional irreducible correp $\rho$, expressed in the basis introduced in Sec.~\ref{sec:RSHam}:
\begin{align}
\rho(\{C_{4z}|000\})&=e^{-i\pi/2J_z} \\
\rho(\{C_{3,111}|000\})&=e^{-2\pi i /(3\sqrt{3})(J_x+J_y+J_z)} \\
\rho(\{T|\half\half\half\})&=\tau_z\sigma_x
\end{align}
Imposing these symmetry constraints, the fourfold fermion at the $R$ point in $P_I432$ takes the particularly simple form
\begin{equation}
H_4(\mathbf{k})=k_z\sigma_z+k_x\sigma_x\tau_x-k_y\sigma_y\tau_x,
\end{equation}
which describes a perfectly isotropic doubled spin-1/2 fermion.\cite{Flicker2018} Note that the twofold degeneracy at generic $\mathbf{k}$ will be lifted by quadratic corrections. Imposing inversion symmetry results in the eight-dimensional co-representation of the little group in $P_Im\bar{3}m$:
\begin{align}
\Delta(\{C_{4z}|000\})&=e^{-i\pi/2J_z}\mu_0 \\
\Delta(\{C_{3,111}|000\})&=e^{-2\pi i /(3\sqrt{3})(J_x+J_y+J_z)}\mu_0 \\
\Delta(\{T|\half\half\half\})&=\tau_z\sigma_x\mu_z\\
\Delta(\{I|000\})&=\mu_x
\end{align}
Imposing these symmetry constraints on the $\mathbf{k}\cdot\mathbf{p}$ Hamiltonian $H(\mathbf{k})$, we find
\begin{equation}
H(\mathbf{k})=H_4(\mathbf{k})\otimes(a\mu_z-b\mu_y).
\end{equation}
This describes a perfectly isotropic double Dirac fermion, which is fourfold degenerate everywhere at linear order. In particular, this Hamiltonian describes two decoupled doubled spin-1/2 fermions of opposite chirality, with chirality matrix $a\mu_z-b\mu_y$. This is distinct from an RS fermion. However, because the doubled spin-1/2 fermions of opposite chirality number are decoupled to linear order, we expect that these double Dirac fermions exhibit an anomalous negative magnetoresistance in analogy to a Dirac semimetal.\cite{burkovchiral}


\section{Chiral threefold fermions}
\label{sec:kp3}

We now turn to the threefold degenerate fermions listed in Table~\ref{table:magneticNF}. Before describing the magnetic groups, we recall the ``spin-1 Weyl'' introduced in Ref.~\onlinecite{Bradlyn2016}, described by the local Hamiltonian,
\begin{equation}
    H_\text{Weyl-1}(\phi,\mathbf{k}) = \begin{pmatrix} 0 & e^{i\phi} k_z & -e^{-i\phi} k_x \\
    e^{-i\phi} k_z & 0 & e^{i\phi} k_y \\
    -e^{i\phi} k_x & e^{-i\phi} k_y & 0 \end{pmatrix}
    \label{eq:Weyl1}
\end{equation}
We have omitted an overall scale factor and energy shift that are not determined by symmetry and which do not impact the topological properties.
At $\phi = \pi/2$, $H_\text{Weyl-1}(\pi/2,\mathbf{k}) = \mathbf{k}\cdot \mathbf{S}$, where $\mathbf{S}_{x,y,z}$ are the generators of $SO(3)$ in the spin-1 representation (this explains the nomenclature).
$H_\text{Weyl-1}$ has the property that at generic values of $\phi$ (i.e., $\phi\neq n\pi/3$), the bands are only degenerate at $\mathbf{k}=\mathbf{0}$. 
Thus, we can evaluate the Chern number of each band over a sphere enclosing the degeneracy point; the Chern number is either $\pm 2$ or $0$ depending on whether the Fermi energy is in the upper/lower bands or the middle band.\cite{Bradlyn2016}

The spin-1 Hamiltonian in Eq.~(\ref{eq:Weyl1}) was derived as the $\mathbf{k}\cdot\mathbf{p}$ Hamiltonian near the $P\equiv (1/4,1/4,1/4)$ point in SG $I2_13$ (199). The little group of this point, which we denote $G^{199}_P$, is generated by $\{ C_{3,111} | \mathbf{0} \}$ and $\{ C_{2z} | \frac{1}{2} 0 \frac{1}{2} \}$,  along with the Bravais lattice translations.\cite{Bilbao2}
The spin-1 representation of this group, as given in the BCS\cite{GroupTheoryPaper}, is
\begin{align}
    \Delta(\{ C_{3,111} | \mathbf{0} \} ) &= \begin{pmatrix} 0 & 0 & -1 \\ 1 & 0 & 0 \\ 0 & 1 & 0 \end{pmatrix} \nonumber\\
    \Delta(\{ C_{2z} | \frac{1}{2} 0 \frac{1}{2} \}) &= {\rm Diag}\left[ 1, 1, -1 \right] 
    \label{eq:199irrep}
\end{align}

Now let us consider the magnetic groups in Table~\ref{table:magneticNF}.

\subsection{Type III groups}
\label{sec:kp3III}
We first consider SGs $P4_3'32'$ $(212.61)$ and $P4_1'32'$ $(213.65)$.
In these groups, the little group at the three-fold degeneracy contains $G^{199}_P$ as a halving subgroup:
\begin{align}
    G_\mathbf{k} &= G^{199}_P \cup  g G^{199}_P,
    \label{eq:halving199}
\end{align}
where $g= \{C_{2,110}|\frac{1}{4}\frac{3}{4}\frac{3}{4}\} \mathcal{T}$ in $P4_3'32'$ and $g= \{C_{2,110}|\frac{3}{4}\frac{1}{4}\frac{1}{4}\} \mathcal{T}$ in $P4_1'32'$.\cite{Gallego2012}
These two groups are enantiomorphic partners: they differ only by one generator, which is indicated by the subscript $4_3$ or $4_1$.

We could derive the representation of $G_\mathbf{k}$ from Eq.~(\ref{eq:199irrep}) with $g$ as an additional generator, as we did in \cref{eq:comm1,eq:comm2,eq:comm3,eq:comm4,eq:comm5,eq:comm6}.
However, this work is already done if we consider the time-reversal invariant supergroup $G_\mathbf{k} \cup \mathcal{T} G_\mathbf{k}$, for which the representations are listed on the BCS.\cite{GroupTheoryPaper}
The time-reversal invariant supergroup of $P4_3'32'$ is $P4_3321' \,(212)$.
Its six-dimensional time-reversal invariant little group representation at $R$, $\Delta^{212}$, is given by,
\begin{align}
    \Delta^{212}(\{ C_{3,111} | \mathbf{0} \} ) &= \Delta(\{ C_{3,111} | \mathbf{0} \} )\otimes \sigma_0\nonumber\\
    \Delta^{212}((\{ C_{2z} | \frac{1}{2} 0 \frac{1}{2} \}) &= \Delta(\{ C_{2z} | \frac{1}{2} 0 \frac{1}{2} \})\otimes \sigma_0 \nonumber\\
    \Delta^{212}(\{C_{2,110} | \frac{1}{4}\frac{3}{4}\frac{3}{4}\}) &= \begin{pmatrix} 0 & i & 0 \\ i & 0 & 0 \\ 0 & 0 & i \end{pmatrix} \otimes \sigma_z \nonumber\\
    \Delta^{212}(\mathcal{T}) &= \mathbb{I}_3 \otimes i\sigma_y \mathcal{K}
    \label{eq:212irrep}
\end{align}
This representation is irreducible.
However, in the magnetic group $P4_3'32'$, $\{C_{2,110} | \frac{1}{4}\frac{3}{4}\frac{3}{4}\}$ and $\mathcal{T}$ are not separately generators, only their product ($g$) is a generator.
Since it is possible to simultaneously block diagonalize $\Delta^{212}(\{ C_{3,111} | \mathbf{0} \} ) ,\, \Delta^{212}(\{ C_{2z} | \frac{1}{2} 0 \frac{1}{2} \})$ and $\Delta^{212}(g)= \Delta^{212}(\{C_{2,110} | \frac{1}{4}\frac{3}{4}\frac{3}{4}\})\Delta^{212}(\mathcal{T})$ into $3\times 3$ blocks, this representation of $P4_3'32'$ is reducible.
Each $3\times 3$ block is itself an irrep of the little co-group of $P4_3'32'$, which constitutes a three-dimensional fermion in the magnetic group $P4_3'32'$.
(The two blocks together comprise a six-fold fermion which is irreducible in the Type II group $P4_3321'$, as was described in Ref.~\onlinecite{Bradlyn2016}.)

Following the same procedure as in Sec.~\ref{sec:RSHam}, the $\mathbf{k}\cdot \mathbf{p}$ Hamiltonian at $R= (\half,\half,\half)$ is exactly given by the spin-1 Weyl Hamiltonian in Eq.~(\ref{eq:Weyl1}); the extra generator $g\mathcal{T}$ does not place any further constraint on the Hamiltonian.
Thus, in $P4_3'32'$, the chiral fermion at the $R$ point will exhibit all of the same exotic features -- for example, double Fermi arcs and a striking Landau level spectrum -- as the three-fold fermion in SG $I2_13$ that we studied extensively in Ref.~\onlinecite{Bradlyn2016}.

In SG $P4_1'32'$, the Type II supergroup is $P4_1321' (213)$.
The equations are identical to Eq.~(\ref{eq:212irrep}) and hence $P4_1'321'$ also hosts a spin-1 Weyl.

We now consider $I4_1'32'\, (214.69)$, which is the body-centered generalization of $P4_3'32'$ and $P4_1'32'$.
It does not have an enantiomorphic partner because it contains both $g=\{ C_{2,110} | \frac{3}{4}\frac{1}{4}\frac{1}{4}\}\mathcal{T}$ and $\{ C_{2,110} | \frac{1}{4}\frac{3}{4}\frac{3}{4}\}\mathcal{T}$ (here and below, when we write $\{ R | \mathbf{t} \}$, the vector $\mathbf{t}$ is written with respect to the conventional cubic lattice vectors, consistent with the notation of the BCS.)
The little group is the same as in the previous two cases, with the addition of the body-centered translation vectors.
Thus, the threefold fermion -- now at the $P$ point -- is identical to that in $I2_13$.

Next we consider $I\bar{4}'3d'\, (220.91)$, which has $g=\{ m_{1\bar{1}0} | \frac{1}{4}\frac{1}{4}\frac{1}{4}\}\mathcal{T}$.\cite{Gallego2012}
In this case, $g$ is not in the little group at $P$; hence, the little group is unitary and thus the low-energy dispersion near $P$ is identical to $G^{199}_P$.
Although in its parent time-reversal invariant space group, $I\bar{4}3d1'$, line nodes emanate from the three-fold degenerate point,\cite{Bradlyn2016}
in the magnetic group the bands are non-degenerate away from the three-fold degeneracy because the glide symmetry that protected the line nodes (in combination with $\{ C_{3,111}|\mathbf{0} \}$) is broken down to $g$.

Finally, we consider the last Type III group with a three-fold degeneracy, $Ia\bar{3}d' \,(230.148)$, which has $g= \{C_{2,110}|\frac{3}{4}\frac{1}{4}\frac{1}{4}\} \mathcal{T}$.\cite{Gallego2012}
In this case, $g$ is in the little group at $P$, and the little group has the same generators as $I4_1'32'$; consequently, it also displays a three-fold degeneracy identical to that in $I2_13$.

\subsection{Type IV groups}
\label{sec:kp3IV}
We now consider the Type IV groups with three-fold degeneracies.
When considering the Type IV groups, we will not use its Type II supergroup because it does not have the same Brillouin zone and hence, generically, the little group in the Type II supergroup contains fewer elements. 
Instead, it turns out to be straightforward to derive the little group representation directly.

We start with SG $P_I2_13 \, (198.11)$, where Eq.~(\ref{eq:halving199}) holds
with $g=\{ E|\frac{1}{2}\frac{1}{2}\frac{1}{2}\}\mathcal{T}$.\cite{Gallego2012}
(As is required for a Type IV group, $P_I2_13$ contains the product of time-reversal and a fractional lattice translation.)
We derive the commutation relations:
\begin{align}
    \{ C_{2z} | \frac{1}{2} 0 \frac{1}{2} \} g &= g \{ C_{2z} | \frac{1}{2} 0 \frac{1}{2} \} \{ E | 110\}\nonumber\\
    \{ C_{3,111} | \mathbf{0} \} g &= g \{ C_{3,111} | \mathbf{0} \}
    \label{eq:198comm}
\end{align}
Acting at the $R$ point ($\mathbf{k} = (\half\half\half ))$, these commutation relations show that $\Delta(\{ C_{2z} | \frac{1}{2} 0 \frac{1}{2} \} )$ and $\Delta(\{ C_{3,111} | \mathbf{0} \})$ both commute with $\Delta(g)$.
Furthermore, since $g^2 = \{E|111\}\mathcal{T}^2$, and $\Delta(\{ E|111\} ) =-1 = \Delta(\mathcal{T}^2)$ (the first equality follows because we are considering the action of translations at $R$ and the second because we are considering double-valued representations), it is consistent to choose $\Delta(g) = \mathcal{K}$, the complex conjugation operator.
The consequence of this additional generator (relative to those of $I2_13$ in Eq.~(\ref{eq:199irrep})) is an extra constraint on the $\mathbf{k} \cdot \mathbf{p}$ Hamiltonian, which pins it to the special exactly solvable point: $H_{\rm Weyl-1}(\pi/2,\mathbf{k})$.
{This Hamiltonian cannot be achieved in the non-magnetic space groups without fine-tuning the parameters.}
To linear order in $\mathbf{k}$, the eigenvalues are exactly equal to $\pm |\mathbf{k}|$ and $0$.
{This is an ideal situation because it maximizes the energy gap between the non-degenerate bands emanating from the three-fold degeneracy and thus allows the maximum range for Fermi arcs.}

The same result is true for the other Type IV groups (SGs $P_I4_332 \, (212.62)$ and $P_I4_132\, (213.66)$): they have yet another additional generator, $\{ C_{2,110} | \frac{3}{4}\frac{1}{4}\frac{1}{4}\}$, on top of the generators for $P_I2_13$, but this generator does not further constrain the $\mathbf{k} \cdot \mathbf{p}$ Hamiltonian.


\section{Sixfold fermions}
\label{sec:kp6}

We now consider the sixfold degenerate fermions in Table~\ref{table:magneticNF}.
In the Type III groups $Pa'3'\, (205.35)$ and $Ia'\bar{3}' \, (206.39)$ the little group at the sixfold degeneracy point is described by Eq.~(\ref{eq:halving199}) with the anti-unitary generator $\{I|\mathbf{0} \}\mathcal{T}$.\cite{Gallego2012} 
We could follow the procedure in Sec.~\ref{sec:kp3III}, but instead we notice that these generators are exactly those of the little group at the $P$ point in the Type II SG $Ia\bar{3}$ (206), which we analyzed in Ref.~\onlinecite{Bradlyn2016}.
The $\mathbf{k} \cdot \mathbf{p}$ Hamiltonian describing the low-energy physics near the $P$ point in $Ia\bar{3}$ is given by (up to an overall scale and constant):
\begin{align}
    H_\text{Dirac-1}&(\phi,r,\theta,\mathbf{k})=\nonumber\\ 
    &\begin{pmatrix} rH_\text{Weyl-1}(\phi,\mathbf{k}) & -ie^{i\theta} H_\text{Weyl-1}(\pi/2,\mathbf{k}) \\ 
    ie^{-i\theta} H_\text{Weyl-1}(\pi/2,\mathbf{k}) & rH_\text{Weyl-1}(-\phi,\mathbf{k}) \end{pmatrix},
    \label{eq:Dirac1}
\end{align}
where $r,\theta$ are real.
We refer to this as a ``spin-1'' Dirac fermion because we showed in Ref.~\onlinecite{Bradlyn2016} that $H_\text{Dirac-1}$ is topologically equivalent to 
$H_\text{Weyl-1}(\phi,\mathbf{k}) \oplus H_\text{Weyl-1}(-\phi,\mathbf{k})$, which is two copies of the spin-1 Weyl Hamiltonian related by the product of time-reversal and inversion symmetry. This is exactly how spin-$\frac{1}{2}$ Dirac fermions are related to spin-$\frac{1}{2}$ Weyl fermions.

In the other two Type III groups, $Ia'\bar{3}'d \,(230.147)$ and $Ia'\bar{3}'d' \,(230.149)$, there is another anti-unitary generator, $\{ C_{2,110} | \frac{3}{4}\frac{1}{4}\frac{1}{4}\}\mathcal{T}$,  in addition to the generators in the previous paragraph\cite{Gallego2012} (note that the generators of the little group can always be chosen such that there is exactly one anti-unitary generator and the rest are unitary; however, for comparison it is more useful in this case to add a second anti-unitary generator).
These generators are exactly those of the little group at the $P$ point in the Type II SG $Ia\bar{3}d$ (230); we showed in Ref.~\onlinecite{Bradlyn2016} that the $\mathbf{k} \cdot \mathbf{p}$ Hamiltonian at that point is also given by Eq.~(\ref{eq:Dirac1}).
Hence, the $\mathbf{k} \cdot \mathbf{p}$ Hamiltonian is identical for all four Type III groups with sixfold fermions (listed in Table~\ref{table:magneticNF}).

There is one Type IV group with a sixfold fermion in Table~\ref{table:magneticNF}, $P_Ia\bar{3}\, (205.36)$.
The little group at the sixfold degeneracy point is generated by $G^{199}_P$, inversion symmetry and the anti-unitary symmetry $\{ E | \frac{1}{2}\frac{1}{2}\frac{1}{2}\}\mathcal{T}$.\cite{Gallego2012}
This group is the little group at the threefold degeneracy point in $P_I2_13$ with the addition of inversion symmetry.
We showed in Sec.~\ref{sec:kp3IV} that a representation for the little group at $R$ in $P_I2_13$ is given by Eq.~(\ref{eq:199irrep}) and $\Delta(\{ E|\half\half\half\} \mathcal{T} ) = \mathcal{K}$.
We now try to find a matrix representative of inversion symmetry, $\{ I | \mathbf{0} \}$.
We will need the commutation relations:
\begin{align}
    \{ C_{2z} | \half 0 \half\} \{ I | \mathbf{0} \} &= \{ I | \mathbf{0} \} \{ C_{2z} |\half 0 \half \} \{ E | \bar{\half} 0 \half\} \nonumber\\
    \{ C_{3,111} | \mathbf{0} \} \{ I | \mathbf{0} \} &= \{ I | \mathbf{0} \} \{ C_{3,111} | \mathbf{0} \} \nonumber\\
    \left( \{ E|\half\half\half\} \mathcal{T} \right) \{ I | \mathbf{0} \} &= \{ I | \mathbf{0} \}  \left( \{ E|\half\half\half\} \mathcal{T} \right) \{E | \bar{\half} \bar{\half} \bar{\half} \} ,
    \label{eq:20536comm}
\end{align}
where $\bar{\half}$ is shorthand for $ -\half$.
At the $R$ point, Eq.~(\ref{eq:20536comm}) translates to the matrix commutation relations:
\begin{align} 
    [ \Delta(\{ C_{2z} | \half 0 \half\}), \Delta(\{ I | \mathbf{0} \} ) ] &= 0 \nonumber\\
    [ \Delta(\{ C_{3,111} | \mathbf{0} \} ), \Delta(\{ I | \mathbf{0} \} ) ] &= 0 \nonumber\\
    \{ \Delta( \{ E|\half\half\half\} \mathcal{T} ), \Delta(\{ I | \mathbf{0} \} ) \} &= 0 
    \label{eq:20536comm2}
\end{align}
In addition, since $\{ I|\mathbf{0} \}^2 = \mathbb{I}$, 
\begin{equation}
    \Delta(\{ I |\mathbf{0} \} )^2 = \mathbb{I}
    \label{eq:20536inv2}
\end{equation}
Using the matrices in Eq.~(\ref{eq:199irrep}) and $\Delta(\{ E|\half\half\half\} \mathcal{T})= \mathcal{K}$, one can check that there is no solution for $\Delta(\{ I | \mathbf{0} \})$ that satisfies Eqs.~(\ref{eq:20536comm2}) and (\ref{eq:20536inv2}).
Hence, the little group does not have a three-dimensional irrep.
Instead, inversion symmetry requires that the original irrep (of the little group at $R$ in $P_I2_13$) doubles in size:
\begin{align}
    \Delta^{P_Ia\bar{3}} (\{ C_{2z} | \half 0 \half \} ) &= \Delta (\{ C_{2z} | \half 0 \half \} ) \otimes \sigma_0 \nonumber\\
    \Delta^{P_Ia\bar{3}} (\{ C_{3,111} | \mathbf{0} \} ) &= \Delta (\{ C_{3,111} | \mathbf{0} \} ) \otimes \sigma_0 \nonumber\\
    \Delta^{P_Ia\bar{3}} (\{ I | \mathbf{0} \} ) &= \mathbb{I}_3 \otimes \sigma_z \nonumber\\
    \Delta^{P_Ia\bar{3}} (\{ E|\half\half\half\} \mathcal{T} ) &= \mathbb{I}_3 \otimes \mathcal{K} \sigma_x
\end{align}
This extra symmetry restricts the $\mathbf{k} \cdot \mathbf{p}$ Hamiltonian to take the special form, 
$H_\text{Dirac-1}(0,0,\theta,\mathbf{k})$, whose spectrum is given by two-fold degenerate bands with energies $\pm |\mathbf{k}|$ and another two-fold degenerate flat band at zero energy.
{As we saw in the Type IV groups with three-fold degeneracies (Sec.~\ref{sec:kp3IV}), the Hamiltonian is pinned to a rather ideal exactly solvable point, which is not achievable in the non-magnetic space groups without fine-tuning.}


\section{Material candidates}
\label{sec:materials}

We present examples of magnetic materials that realize the nodal fermions we have described.
In particular, we introduce a table of Wyckoff positions (Table~\ref{tab:magneticWyckoff}) that are compatible with having a magnetic moment.
This allows us to find magnetic materials by searching for compounds in a crystal structure database (for example, the ICSD) that have a magnetic element in one of the Wyckoff positions in Table~\ref{tab:magneticWyckoff}.
The examples we present serve as a proof-of-principle that this method works to find magnetic multifold fermions.
We postpone a more thorough search to future work.

\subsection{Non-magnetic}

In Ref.~\onlinecite{Bradlyn2016}, we presented the materials Ta$_3$Sb, LaPd$_3$S$_4$ and Nb$_3$Bi in SG $Pm\bar{3}n$.
Although it was not realized in that work, these compounds present solid state realizations of the RS fermion.

\subsection{Magnetic Type I groups}

Table~\ref{tab:NF} lists the Type II space groups that can host 3-, 6- and 8-fold fermions; there we noted that 3-fold fermions (and not 6- and 8-) can also appear in Type I groups, which have no anti-unitary symmetries.

\begin{figure*}[t]
\includegraphics[width=0.6\textwidth,angle=-90]{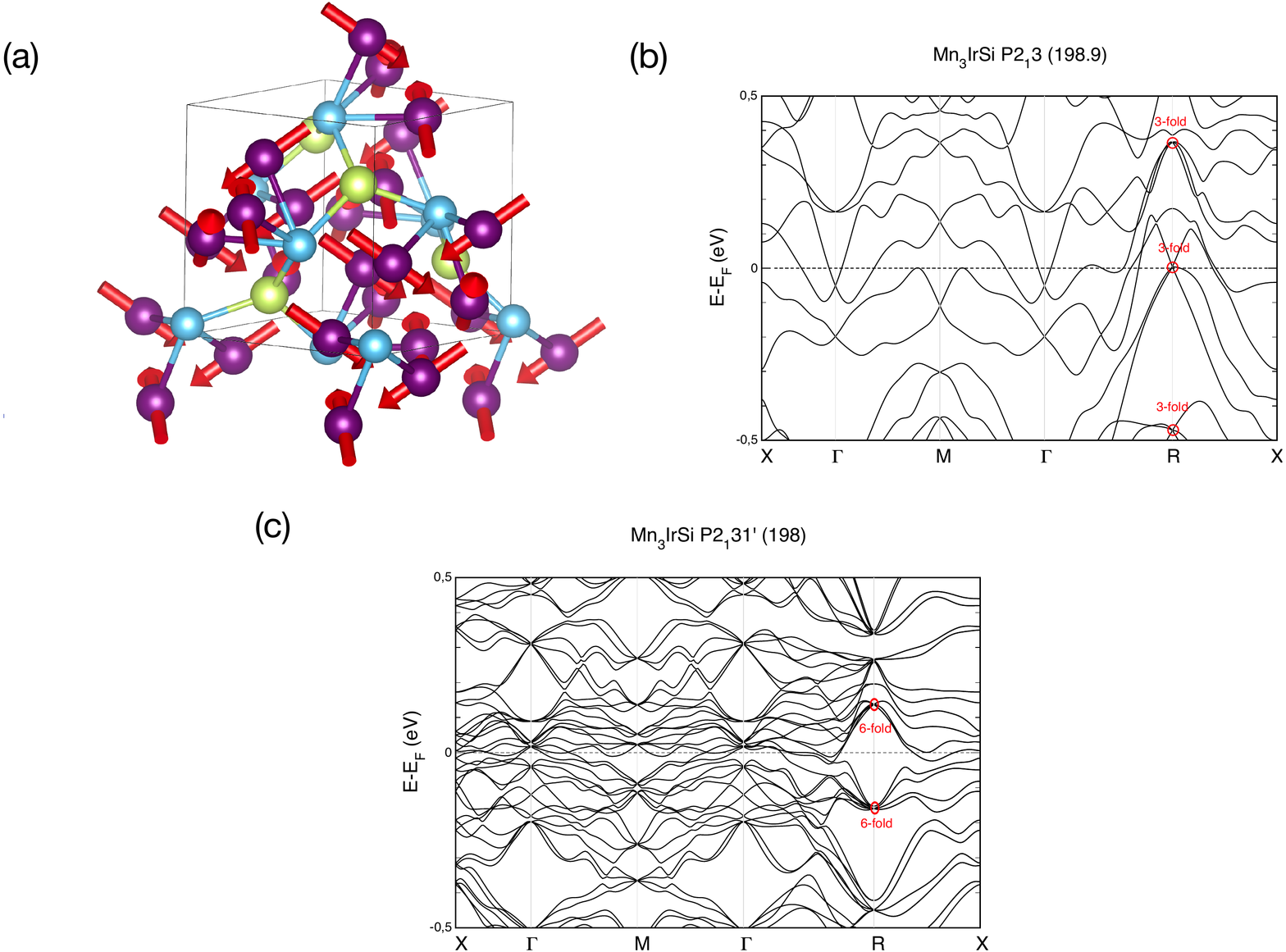}
\caption{(a) Crystal structure of Mn$_3$IrSi in $P2_13$. It forms a noncollinear antiferromagnetic structure with Mn magnetic moments found on three dimensional networks of corner linked triangles. Band structures in $P2_13$ (magnetically ordered) and $P2_131'$ (TRS) with 3- and 6-fold fermions are indicated by red circles in (b) and (c) respectively.}\label{fig:TypeI}
\end{figure*}

One example is realized by Mn$_3$IrSi in SG $P2_13$ (198). 
In this compound the magnetic atom (Mn) resides in the $12b$ Wyckoff position. Measurements in Ref.~\onlinecite{ERIKSSON2004823,Eriksson04Mn3IrSi} revealed a complex magnetic order that did not change the size of the unit cell; the corresponding theory further showed that the magnetic order in the lowest energy configuration preserved all of the unitary symmetry operations (see Fig.~\ref{fig:TypeI} (a)). Thus, the reported magnetic order is described by the Type I group $P2_13$, which appears in Table~\ref{tab:NF} as having a 3-fold(6-fold) degeneracy when time-reversal symmetry is broken(present). We performed spin polarized calculations with a Hubbard  U chosen to be 3 eV (a common value for 3d transition metals) to reveal these three-fold/six-fold degeneracies in Fig.~\ref{fig:TypeI} (b)-(c). 
Absent any Mott physics, we expect the three-fold(six-fold) degeneracies to be present below(above) the N{\'e}el temperature. Other compounds of the same family are Mn$_3$IrGe, Mn$_3$Ir$_{1{-}y}$Co$_y$Si and Mn$_3$CoSi$_{1{-}x}$Ge$_x$\cite{ERIKSSON2004823,Eriksson04Mn3IrSi}.

{SG 198 with time-reversal symmetry hosts a chiral \textit{six-fold} fermion, which is possible because the crystal belongs to one of the original Sohncke space groups, which contain only rotational symmetries.\cite{Chang2017,Flicker2018}}
Time-reversal symmetry causes two copies of the three-dimensional chiral irrep $\bar{R}_7$ to be degenerate in energy.
Below the N{\'e}el temperature, the two copies split.
In both cases, the physical consequences of a chiral fermion should be observable, in particular, quantization of the circular photogalvanic effect \cite{de2017quantized,Flicker2018,Chang2017,rees2019quantized} and large Fermi arcs\cite{Schroter2018,Chang2017}.

\subsection{Magnetic Type III and IV groups}

We now turn to the challenge of finding materials in the magnetic space groups.
As mentioned earlier, this is a nontrivial task because there does not exist a comprehensive database of magnetic materials sorted by their magnetic space group (although some can be found on the BCS\cite{MagneticBCS1,MagneticBCS2}) and in many cases, the magnetic space group is never reported. 
Thus, most generally, to find material candidates in the magnetic space group $G$, we search the ICSD\cite{ICSD} by the nonmagnetic space group (denoted $G'$ in Table~\ref{table:magneticNF})
for materials with magnetic elements.

Even without knowing the magnetic order, some candidate materials can be ruled out because they do not have any atom in a Wyckoff position compatible with a magnetic moment.
(Not all Wyckoff positions are compatible: for example, if $G$ contains $C_{2x}, C_{2y}$ and $C_{2z}$, a site invariant under all three rotations is incompatible with an ordered magnetic spin.)
The compatible Wyckoff positions are listed in Table~\ref{tab:magneticWyckoff}; they were obtained via the MWYCKPOS\cite{Gallego2012,PerezMato2015} tool on the BCS.
(Note that the magnetic Wyckoff positions are not always the same as the Wyckoff position in the nonmagnetic space group, and might not even have the same multiplicity.)

The complexity of the magnetic structures escalates when we look for multifold fermions in Type III and IV groups. We did not find in the literature magnetic materials displaying these phases; however, we deduced compounds in time-reversal invariant parent SGs that could potentially host these magnetic structures, and we have computed the magnetic space group that minimizes the energy from ab initio calculation. We hope that the proof-of-principle materials presented here serve as inspiration for experimentalists and crystallographers to more systematically classify magnetic structures by symmetry.

\begin{figure*}[t]
\vspace*{-10mm}
\includegraphics[width=0.6\textwidth,angle=-90]{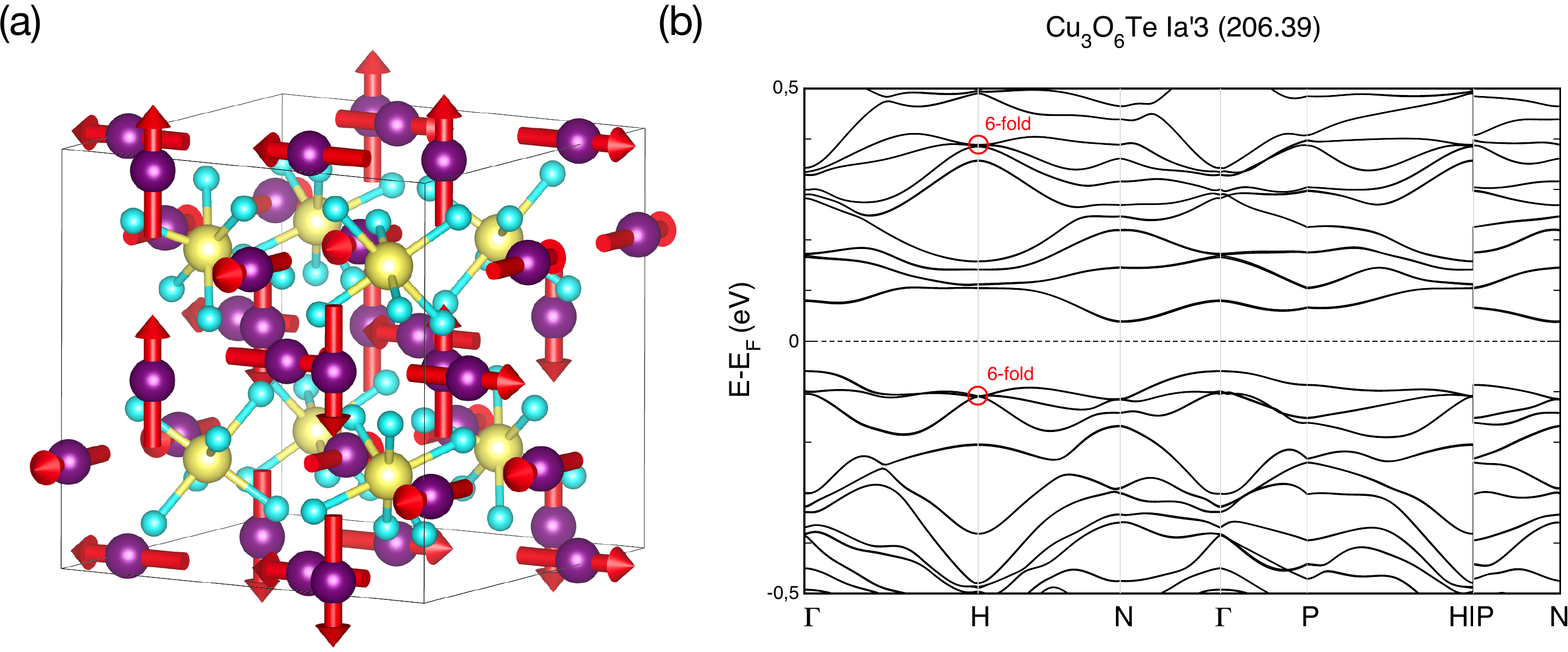}
\vspace*{-25mm}
\caption{(a) Crystal structure of Cu$_3$O$_6$Te in $Ia'3$. (b) Band structure along high symmetry points with 6-fold degeneracies indicated by a red circle.}\label{TypeIII}
\end{figure*}

\begin{figure*}[t]
\includegraphics[width=0.6\textwidth]{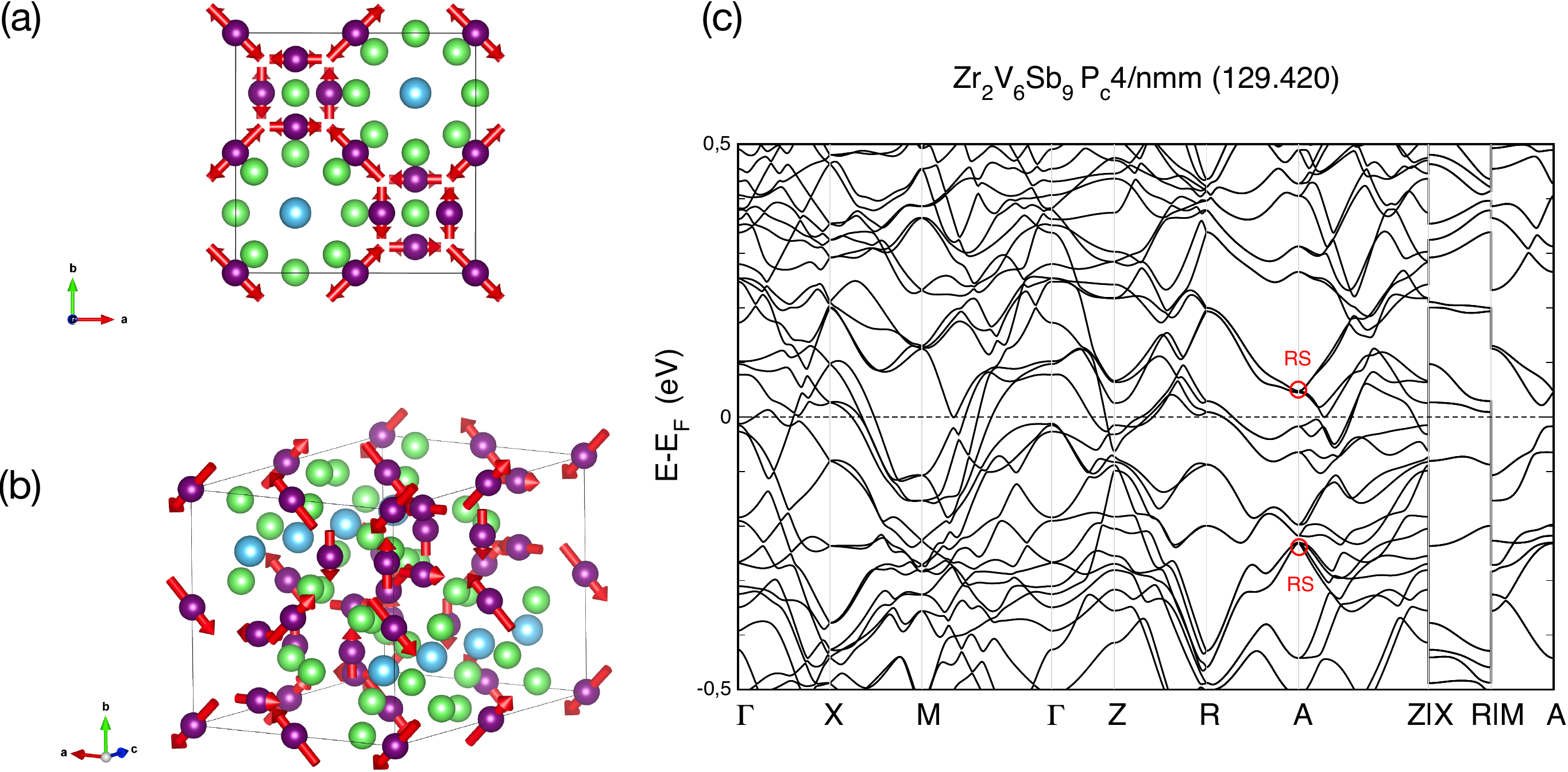}
\caption{(a)-(b) In-plane and general view of the magnetic structure of Zr$_2$V$_6$Sb$_9$ in $P_C4/nmm$ 
respectively. (c) Band structure along high symmetry points; Rarita Swinger fermions are surrounded by red circles.}\label{fig:TypeIV}
\end{figure*}

Our Type III candidate is Cu$_3$O$_6$Te\cite{Falcka15888} in $Ia'$3, which consists of distorted close-packed O layers in c stacking and Cu and Te in octahedral voids (see Fig.\ref{TypeIII}(a)). The magnetic atoms (Cu) reside in the $24d$ Wyckoff position. 
This compound was reported to have a long-range magnetic order\cite{MANSSON2012142} below 61.7 K, although the magnetic structure was undetermined.
Recently, using inelastic neutron scattering, a long-range collinear antiferromagnetic order below the transition T$_N$ temperature of 61K was reported, with spins aligned along [111] direction.\cite{PMID29968725} 
This is a different configuration than in our computational experiment; although the magnetic phase we have computed has not been reported, different growth conditions or external parameters could tune the material into it.

In the Type IV groups, we report the antimonide Zr$_2$V$_6$Sb$_9$, in $P_C4/nmm$, 
to be a candidate material to display RS fermions\cite{Sb9V6Zr2}. It contains chains of face-sharing square Zr-centered Sb$_8$ antiprisms, which are surrounded by V layers, where the V atoms are situated in severely distorted Sb$_6$ octahedra, as shown in Fig. \ref{fig:TypeIV} (a) and (b). 
The magnetism comes from antiferromagnetic ordering on the V atoms, which reside on two different Wyckoff positions: the $4d$ and $8i$ positions in the Type II parent space group $P4/nmm1'$, which correspond to the $8d$ and $16i$ positions in $P_C4/nmm$.
Generically, the magnitude of the magnetic moment is different on different Wyckoff positions and it is not necessary for both positions to have a non-zero moment in order to display RS fermions.
The material was reported to be paramagnetic at low temperature, however attempts to synthesize crystals suitable for direction dependent measurements were unsuccessful. Fig. \ref{fig:TypeIV} (c) shows the band structure of Zr$_2$V$_6$Sb$_9$, with RS fermions surrounded by red circles.

\begin{table}[t]
    \centering
    \begin{tabular}{cccc}
        Deg. & Type & BNS & Wyckoff\\
        \hline
        3-fold & III & 212.61 & all\\
        && 213.65 & all\\
        && 214.69 & all\\
        && 220.91 & $c,d,e$\\
        && 230.148 & $a,b,c,e,f,g,h$\\
        \hline
        3-fold & IV & 198.11 & all\\
        && 212.62 & $b,c,d,f,g,h,i$\\
        && 213.66 & $a,c,d,e,f,g,h,i$\\
        \hline
        6-fold & III & 205.35 & $c,d$\\
        && 206.39 & $c,d,e$\\
        && 230.147 & $b,c,d,e,f,g,h$\\
        && 230.149 & $e,f,g,h$\\
        \hline
        6-fold & IV & 205.36 & $b,c,d,e$\\
        \hline
        8-fold & III & 222.102 & $e,f,g,h,i$\\
        && 223.108 & $f,g,h,i,j,k,l$\\
        && 230.149 & $e,f,g,h$\\
        \hline
        8-fold & IV & 125.374 & $e,f,h,i,j,k,l,m$\\
        && 126.385 & $e,f,h,i,j,k,l,m,n$\\
        && 129.420 & $d,g,h,i,j,k$\\
        && 131.445 & $e,k,l,m,n,o,p$\\
        && 132.458 & $i,j,k,l$\\
        && 136.504 & $h,i,j,k,l$\\
        && 215.73 & $f,g,h$\\
        && 221.97 & $i,j,k,l$\\
        && 224.115 & $g,h,i,j,k,l$
    \end{tabular}
    \caption{Magnetic Wyckoff positions compatible with a non-zero magnetic moment. The first column gives the degeneracy, the second column the type of magnetic space group, the third column the BNS number of the group and the fourth column gives the Wyckoff positions that can host a non-zero magnetic moment.}
    \label{tab:magneticWyckoff}
\end{table}


\section{Discussion}
\label{sec:discussion}

In this paper, we have exhaustively enumerated the unconventional fermionic quasiparticles that can occur in magnetic metals and semimetals, shown in Table~\ref{table:magneticNF}.
Additionally, we have explored the properties of these highly degenerate excitations using the $\mathbf{k}\cdot\mathbf{p}$ expansion. In doing so we have highlighted that many eightfold degenerate fermions in both the magnetic and nonmagnetic space groups realize a generalization of the massless Rarita-Schwinger Hamiltonian, in many cases with an approximately conserved chirality operator. 
{In addition, we showed that the Type IV magnetic groups contain multifold fermions that reside at idealized exactly solvable points in the phase diagram that are not achievable in non-magnetic space groups without fine-tuning.}
Finally, we have presented several material candidates which can serve as a platform for finding multifold nodal fermions in realistic magnetic systems.

This work is not the end of the story, but rather can serve as a foundation to explore several future directions. First, we expect by analogy with the nonmagnetic case that the magnetic multifold fermions presented here play a crucial role in determining the minimal insulating filling of magnetic systems.\cite{WatanabeMagnetic} Understanding the insulating filling constraints in magnetic systems will be crucial for finding magnetic multifold nodal points in real materials, since in a strongly interacting magnet the notion of a quasiparticle band can in general only be well-defined near the Fermi energy.\cite{schriefferspectral} We thus must look for materials with half-filled groups of bands near the Fermi energy in order to resolve these multifold magnetic fermions.

Second, we expect these fermions to exhibit a host of exotic topological phenomena, in analogy with their nonmagnetic counterparts. Chiral threefold and sixfold fermions will host surface Fermi arcs; since time-reversal symmetry is broken, these arcs need not come in pairs. This opens up the possibility to predict and observe large anomalous Hall conductances, as well as novel nonlinear optical response.\cite{Flicker2018} Additionally, even the nonchiral magnetic degeneracis (such as the RS fermions) may exhibit exotic Landau level spectra and anomalous magnetoresistance due to an approximately conserved chirality.\cite{burkovchiral}

Finally, we expect that these multifold fermions can serve as a platform for field-tunable topological behavior. Previous studies\cite{Cano2017,Wang2016} have shown how magnetic fields can be used to generate Weyl fermions from topologically trivial band structures. From our analysis here, this approach to designer topological semimetals can be extended to higher-Chern number multifold semimetals. Furthermore, we expect that magnetic transitions which gap these unconventional fermions could lead to exotic magnetic (higher-order) topological phases. 

Using the tools we have developed here, we hope to inspire further work along these directions. In particular, we hope to highlight the benefits that could be reaped from a more systematic symmetry classification of reported magnetic structures in materials databases.


\section{Acknowledgements}

We are grateful for the hospitality of the Aspen Center for Physics where part of this work was performed. 
The Aspen Center for Physics is supported by National Science Foundation grant PHY-1607611.
We also benefited from conversations with M.~I.~Aroyo, B.~A.~Bernevig, Luis Elcoro and Benjamin Wieder. Additionally, BB acknowledges fruitful discussions with M.~Stone.
JC acknowledges support from the Flatiron Institute, a division of the Simons Foundation. MGV  acknowledges the IS2016-75862-P national project of the Spanish MINECO.


\appendix

\section{Proof of Eq.~(\ref{eq:TypeIIIanti})}
\label{sec:prooflittlegroup}
Given a magnetic group, $G$, and a momentum, $\mathbf{k}$, 
if the little group $G_\mathbf{k}$ contains anti-unitary elements, then there must exist an $h_0\in H$ that satisfies Eq.~(\ref{eq:defh0}), which we have repeated here for convenience:
\begin{equation}
\mathcal{T} g_0 h_0 \mathbf{k} = \mathbf{k}.
\tag{\ref{eq:defh0}}
\end{equation}
(For if there was no such $h_0$, then $G_\mathbf{k}$ would be unitary.)

Here we prove that the little group $G_\mathbf{k}$ is given by:
\begin{equation}
G_\mathbf{k} = H_\mathbf{k} \cup \mathcal{T} g_0 h_0 H_\mathbf{k}.
\tag{\ref{eq:TypeIIIanti}}
\end{equation}

Proof: first, since $H_\mathbf{k}$ is the unitary part of $G_\mathbf{k}$, we need only prove that the set $\mathcal{T} g_0 h_0 H_\mathbf{k}$ is exactly the anti-unitary part of $G_\mathbf{k}$.
In the first direction, $\mathcal{T} g_0 h_0 H_\mathbf{k}\subset G_\mathbf{k}$ because for each $h_\mathbf{k} \in H_\mathbf{k}$, 
\begin{equation}
\mathcal{T} g_0h_0 h_\mathbf{k} \mathbf{k} = \mathcal{T} g_0h_0\mathbf{k} = \mathbf{k},
\end{equation}
where the first equality follows because $h_\mathbf{k} \in H_\mathbf{k}$ and the second equality follows from Eq.~(\ref{eq:defh0}).
In the other direction, if $g\in G_\mathbf{k}$ is anti-unitary, then $g=\mathcal{T}g_0 h$ for some $h\in H$; thus, $\mathcal{T}g_0 h\mathbf{k} = \mathbf{k}$.
Combined with Eq.~(\ref{eq:defh0}), $h\mathbf{k} = h_0\mathbf{k}$. 
Hence, $h\in h_0 H_\mathbf{k}$, which completes the proof.

\section{Little groups with 3- or 4- dimensional double-valued (spinful) irreps}
\label{sec:listdoubleirreps}

We have checked (using the BANDREP application on the BCS server\cite{GroupTheoryPaper}) that the largest double-valued (spinful) irreps of little groups that appear in the unitary Type I space groups are 3- and 4- dimensional. 
These are listed in Tables~\ref{table:3ddouble} and~\ref{table:4ddouble}, respectively.
We have listed the largest spinless (single-valued) irreps of little groups in Appendix~\ref{sec:listsingleirreps}.

Notice that all of the little groups with 3-dimensional irreps (listed in Table~\ref{table:3ddouble}) appear in Table I of Ref~\onlinecite{Bradlyn2016} as 3- or 6-dimensional irreps because in Ref~\onlinecite{Bradlyn2016} time reversal symmetry is included, which can either double the dimension of the irrep or leave it unchanged.
All of the 8-dimensional irreps in Table I of Ref~\onlinecite{Bradlyn2016} have doubled with time reversal symmetry and hence appear here in Table~\ref{table:4ddouble}. However, there are some 4-dimensional irreps in Table~\ref{table:4ddouble} that do not appear in Table 1 of Ref~\onlinecite{Bradlyn2016}; these are exactly the irreps that do not double in the presence of time reversal symmetry.

\begin{table}[t]
\centering
\begin{tabular}{ccc}
SG & $\mathbf{k}$ & Irrep\\
\hline
198 & $R$ & $\bar{R}_7$\\
\hline
199 & $P$ & $\bar{P}_7$\\
\hline
205 & $R$ & $\bar{R}_{10}, \bar{R}_{11}$\\
\hline
206, 214 & $P$ & $\bar{P}_7$ \\
\hline
212, 213 & $R$ & $\bar{R}_7, \bar{R}_8$\\
\hline
220, 230 & $P$ & $\bar{P}_7, \bar{P}_8$\\
\end{tabular}
\caption{Unitary (Type 1) space groups that host 3-dimensional double-valued irreps. The first column gives the space group number, the second column the $\mathbf{k}$ at which the irrep exists and the third column gives the name(s) of the irrep(s), using the notation of the BANDREP app on the BCS.}
\label{table:3ddouble}
\end{table}

\begin{table}[t]
\centering
\begin{tabular}{ccc}
SG & $\mathbf{k}$ & Irrep\\
\hline
125, 126, 129, 130 & $A$ & $\bar{A}_5$\\
 & $M$ & $\bar{M}_5$\\
\hline
131, 132, 135, 136 & $A$ & $\bar{A}_5$\\
 & $Z$ & $\bar{Z}_5$\\
\hline
133, 134, 137, 138 & $M$ & $\bar{M}_5$\\
 & $Z$ & $\bar{Z}_5$\\
 \hline
 141, 142 & $M$ & $\bar{M}_5$\\
 \hline
 193, 194 & $A$ & $\bar{A}_6$\\
\hline
207, 208, 215, 218 & $\Gamma$ &$\bar{\Gamma}_8$\\
& $R$ & $\bar{R}_8$ \\
\hline
209, 210, 212, 213, 216, 219 & $\Gamma$ &$\bar{\Gamma}_8$\\
\hline
211,214, 220 & $\Gamma$ & $\bar{\Gamma}_8$\\
& $H$ & $\bar{H}_8$ \\
\hline
217 & $\Gamma$ & $\bar{\Gamma}_8$\\
& $H$ & $\bar{H}_8$ \\
& $P$ & $\bar{P}_8$\\
\hline
221 & $\Gamma$ & $\bar{\Gamma}_{10}, \bar{\Gamma}_{11}$\\
& $R$ & $\bar{R}_{10}, \bar{R}_{11}$ \\
\hline
222 & $\Gamma$ & $\bar{\Gamma}_{10}, \bar{\Gamma}_{11}$\\
& $M$ & $\bar{M}_5$\\
& $R$ & $\bar{R}_5, \bar{R}_6, \bar{R}_7$\\
\hline 
223  & $\Gamma$ & $\bar{\Gamma}_{10}, \bar{\Gamma}_{11}$\\
& $R$ & $\bar{R}_5, \bar{R}_6, \bar{R}_7$\\
& $X$ & $\bar{X}_5$\\
\hline
224 & $\Gamma$ & $\bar{\Gamma}_{10}, \bar{\Gamma}_{11}$\\
& $M$ & $\bar{M}_5$\\
& $R$ & $\bar{R}_{10}, \bar{R}_{11}$\\
& $X$ & $\bar{X}_5$\\
\hline
225, 226 & $\Gamma$ & $\bar{\Gamma}_{10}, \bar{\Gamma}_{11}$\\
\hline
227, 228 & $\Gamma$ & $\bar{\Gamma}_{10}, \bar{\Gamma}_{11}$\\
& $X$ & $\bar{X}_5$\\
\hline
229 & $\Gamma$ & $\bar{\Gamma}_{10}, \bar{\Gamma}_{11}$\\
& $H$ & $\bar{H}_{10}, \bar{H}_{11}$\\
& $P$ & $\bar{P}_8$\\
\hline
230 & $\Gamma$ & $\bar{\Gamma}_{10}, \bar{\Gamma}_{11}$\\
& $H$ & $\bar{H}_4, \bar{H}_5, \bar{H}_6, \bar{H}_7$ 
\end{tabular}
\caption{Unitary (Type 1) space groups that host 4-dimensional double-valued irreps. The first column gives the space group number, the second column the $\mathbf{k}$ at which the irrep exists and the third column gives the name of the irrep, using the notation of the BANDREP app on the BCS.}
\label{table:4ddouble}
\end{table}

\section{Little groups with 3-, 4- or 6- dimensional single-valued (spinless) irreps}
\label{sec:listsingleirreps}

We have checked (using the BANDREP application on the BCS server) that the largest single-valued (spinless) irreps of little groups that appear in the unitary (Type I) space groups are 3-, 4- and 6- dimensional. 
These are listed in Tables~\ref{table:3dsingle}, ~\ref{table:4dsingle}, and \ref{table:6dsingle}, respectively.

\begin{table}[t]
\centering
\begin{tabular}{ccc}
SG & $\mathbf{k}$ & Irrep\\
\hline
195 & $\Gamma$ & $\Gamma_4$\\
& $R$ & $R_4$ \\
\hline
196, 198 & $\Gamma$ & $\Gamma_4$\\
\hline
197 & $\Gamma$ &$\Gamma_4$\\
& $H$ & $H_4$\\
& $P$ & $P_4$\\
\hline
199 & $\Gamma$ &$\Gamma_4$\\
& $H$ & $H_4$\\
\hline
200, 201 & $\Gamma$ & $\Gamma_4^+, \Gamma_4^-$\\
& $R$ & $R_4^+, R_4^-$\\
\hline
202, 203, 205 & $\Gamma$ & $\Gamma_4^+, \Gamma_4^-$\\
\hline
204 & $\Gamma$ & $\Gamma_4^+, \Gamma_4^-$\\
& $H$ & $H_4^+, H_4^-$ \\
& $P$ & $P_4$\\
\hline
206 & $\Gamma$ & $\Gamma_4^+, \Gamma_4^-$\\
& $H$ & $H_4^+, H_4^-$ \\
\hline
207, 208, 215, 218 & $\Gamma$ & $\Gamma_4, \Gamma_5$\\
& $R$ & $R_4, R_5$ \\
\hline
209, 210, 219 & $\Gamma$ & $\Gamma_4, \Gamma_5$\\
\hline
211 & $\Gamma$ &$\Gamma_4, \Gamma_5$\\
& $H$ & $H_4, H_5$\\
& $P$ & $P_4$\\
\hline
212, 213, 216 & $\Gamma$ &$\Gamma_4, \Gamma_5$\\
\hline
214, 220 & $\Gamma$ &$\Gamma_4, \Gamma_5$\\
& $H$ & $H_4, H_5$\\
\hline
217 & $\Gamma$ &$\Gamma_4, \Gamma_5$\\
& $H$ & $H_4, H_5$\\
& $P$ & $P_4, P_5$\\
\hline
221, 224 & $\Gamma$ & $\Gamma_4^+, \Gamma_4^-,\Gamma_5^+, \Gamma_5^-$\\
& $R$ & $R_4^+, R_4^-,R_5^+, R_5^-$\\
\hline
222, 223, 225, 226, 227, 228, 230 & $\Gamma$ & $\Gamma_4^+, \Gamma_4^-,\Gamma_5^+, \Gamma_5^-$\\
\hline
229 & $\Gamma$ & $\Gamma_4^+, \Gamma_4^-,\Gamma_5^+, \Gamma_5^-$\\
& $H$ & $H_4^+, H_4^-, H_5^+, H_5^-$\\
& $P$ & $P_4, P_5$
\end{tabular}
\caption{Unitary (Type 1) space groups that host 3-dimensional single-valued irreps. The first column gives the space group number, the second column the $\mathbf{k}$ at which the irrep exists and the third column gives the name(s) of the irrep(s), using the notation of the BANDREP app on the BCS.}
\label{table:3dsingle}
\end{table}

\begin{table}[t]
\centering
\begin{tabular}{ccc}
SG & $\mathbf{k}$ & Irrep\\
\hline
193, 194 & $A$ & $A_3$ \\
\hline
212, 213 & $R$ & $R_3$\\
\hline
220, 230 & $P$ & $P_3$\\
\end{tabular}
\caption{Unitary (Type 1) space groups that host 4-dimensional single-valued irreps. The first column gives the space group number, the second column the $\mathbf{k}$ at which the irrep exists and the third column gives the name of the irrep, using the notation of the BANDREP app on the BCS.}
\label{table:4dsingle}
\end{table}

\begin{table}[t]
\centering
\begin{tabular}{ccc}
SG & $\mathbf{k}$ & Irrep\\
\hline
222, 223 & $R$ & $R_4$\\
230 & $H$ & $H_4$
\end{tabular}
\caption{Unitary (Type 1) space groups that host 6-dimensional single-valued irreps. The first column gives the space group number, the second column the $\mathbf{k}$ at which the irrep exists and the third column gives the name(s) of the irrep(s), using the notation of the BANDREP app on the BCS.}
\label{table:6dsingle}
\end{table}

\bibliography{MagneticNF}
\end{document}